\documentclass[10pt, two column, twoside]{IEEEtran}
\pdfoutput=1

\usepackage{amssymb}
\usepackage{amsmath}
\usepackage{bm}
\usepackage[lined,boxed,commentsnumbered, ruled]{algorithm2e}
\usepackage{mathrsfs}
\usepackage{algorithmic}
\usepackage{tikz}
\usetikzlibrary{arrows}
\usepackage{subfigure}
\usepackage{graphicx,booktabs,multirow}
\usepackage{url}

\definecolor{colorhkust}{RGB}{20,43,140}
\definecolor{colortsinghua}{RGB}{116,52,129}
\definecolor{color1}{RGB}{128,0,0}

\newtheorem{lemma}{Lemma}
\newtheorem{theorem}{Theorem}

\newtheorem{proposition}{Proposition}
\newtheorem{definition}{Definition}

\newcommand{\A}{\mathcal{A}}
\newcommand{\N}{\mathcal{N}}
\newcommand{\K}{\mathcal{K}}
\newcommand{\argmin}{\operatornamewithlimits{arg\,min}}
\newcommand{\RNum}[1]{\lowercase\expandafter{\romannumeral #1\relax}}
\newcommand{\la}{\langle}
\newcommand{\ra}{\rangle}

\begin{document}

\title{Sparse Optimization for Green Edge AI Inference}

\author{Xiangyu~Yang,~\IEEEmembership{Student~Member,~IEEE,}
 Sheng~Hua,~\IEEEmembership{Student~Member,~IEEE,}
  Yuanming~Shi{*},~\IEEEmembership{Member,~IEEE,}
 ~Hao~Wang,~\IEEEmembership{Member,~IEEE,}
  ~Jun~Zhang,~\IEEEmembership{Senior Member,~IEEE,}
   and Khaled B. Letaief,~\IEEEmembership{Fellow,~IEEE,}
 \thanks{X.~Yang and S.~Hua are with the School of Information Science and Technology, ShanghaiTech University, Shanghai~201210, China,  also with the Shanghai Institute of Microsystem and Information Technology, Chinese Academy of Sciences, Shanghai 200050, China, and also with the University of Chinese Academy of Sciences, Beijing 100049, China.  (e-mail: \{yangxy3, huasheng\}@shanghaitech.edu.cn).}
 \thanks{Y. Shi and H.~Wang are with the School of Information Science and Technology,
ShanghaiTech University, Shanghai, China (e-mail: \{shiym, wanghao1\}@shanghaitech.edu.cn).}
 \thanks{J. Zhang is with the Department of Electronic and Information Engi- neering, The Hong Kong Polytechnic University, Hong Kong (e-mail: jun-eie.zhang@polyu.edu.hk).}
 \thanks{K. B. Letaief is with the Department of Electronic and Computer Engineer- ing, Hong Kong University of Science and Technology, Hong Kong (e-mail: eekhaled@ust.hk). He is also with Peng Cheng Laboratory, Shenzhen, China.}
 \thanks{{*}(\textit{Corresponding author: Yuanming~Shi.})}
 \thanks{Part of this work was presented at the IEEE 90th Vehicular Technology Conference (VTC2019-Fall), Honolulu, Hawaii, USA, Sept. 2019~\cite{hua2019deep}.}
}

\maketitle

\begin{abstract}
With the rapid upsurge of deep learning tasks at the network edge, effective edge artificial intelligence (AI) inference becomes critical to provide low-latency intelligent services for mobile users via leveraging the edge computing capability. In such scenarios, energy efficiency becomes a primary concern. In this paper, we present a joint inference task selection and downlink beamforming strategy to achieve energy-efficient edge AI inference through minimizing the overall power consumption consisting of both computation and transmission power consumption, yielding a mixed combinatorial optimization problem. By exploiting the inherent connections between the set of task selection and group sparsity structural transmit beamforming vector, we reformulate the optimization as a group sparse beamforming problem. To solve this challenging problem, we propose a log-sum function based three-stage approach. By adopting the log-sum function to enhance the group sparsity, a proximal iteratively reweighted algorithm is developed. Furthermore, we establish the global convergence analysis and provide the ergodic worst-case convergence rate for this algorithm. Simulation results will demonstrate the effectiveness of the proposed approach for improving energy efficiency in edge AI inference systems.

\end{abstract}


\section{Introduction}
\IEEEPARstart{T}{HE} availability of big data and computing power, along with the advances in the optimization algorithms, has triggered a booming era of artificial intelligence (AI). Notably, deep learning~\cite{lecun2015deep} is regarded as the most popular sector in modern AI and has achieved exciting breakthroughs in applications such as speech recognition, computer vision~\cite{voulodimos2018deep}, etc. Benefiting from these achievements, AI is becoming a promising tool that streamlines people's decision-making process and facilitates the development of diversified intelligence services (e.g., virtual personal assistant, recommendation system, etc). Meanwhile, with the proliferation of mobile computing as well as Internet-of-Things (IoT) devices, massive real-time data are generated locally~\cite{index2019global}. However, it is widely acknowledged that traditional cloud-based computing~\cite{zhang2010cloud,zhisheng2016cloud} faces challenges (e.g., latency, privacy and network congestion) for supporting the ubiquitous AI-empowered applications on mobile devices~\cite{chen2019deep}.

In contrast, edge AI is a promising approach, which can tackle the above concerns, via fusing mobile edge computing~\cite{mao2017survey} with AI-enabled techniques (e.g., deep neural networks (DNNs)). By pushing AI models to the network edge, it brings the edge servers close to the requesting mobile devices and thus enables low-latency and privacy-preserving. Notably, edge AI is envisioned as the key ingredient of future intelligent $6$G networks~\cite{letaief2019roadmap,zhang20196g,zhang2020mobile,huang2019overview}, which fully unleashes the potentials for mobile communication and computation. Typically, edge AI consists of two phases of edge training and edge inference. As
for the edge training phase, the training of AI models can be
performed on cloud, edge or end devices~\cite{yang2020federated}, however, this is
beyond the scope of this work. By deploying trained AI models and implementing model inference at network edge, this paper mainly focuses on edge inference. Following~\cite{chen2019deep,zhou2019edge}, the edge AI inference architecture is generally classified into three major types: 
\begin{itemize}
        \item \textbf{On-device inference:} It performs the model inference directly on end devices where DNN models are deployed. While some enabling techniques (e.g., model compression~\cite{jiang2019layer,ren2019admm}, hardware speedup~\cite{sze2017efficient}) have been proposed to facilitate the deployment of the DNN model, it still poses challenges for resource-limited (e.g., memory, power budget and computation) end devices~\cite{plastiras2018edge}. To mitigate such concerns, on-device distributed computing is envisioned as a promising solution for on-device inference, which enables AI model inference across multiple distributed end devices~\cite{yang2019data}. 

        \item \textbf{Joint device-edge inference:} This mode carries out the AI model inference in a device-edge cooperation fashion~\cite{chen2019deep} with the model partition and model early-exit techniques~\cite{li2019edge}. While device-edge cooperation is flexible and enables low correspondence-latency edge inference, it may still have high resource requirements for end devices due to the resource-demanding nature of DNNs~\cite{he2016deep}.  

        \item \textbf{Edge server inference:} Such methods transfer the raw input data to edge serves for processing, which then return inference results to  end-users~\cite{hua2019reconfigurable,yang2019energy}. Edge server inference is particularly suitable for those computation-intensive tasks. Nonetheless, the inference performance relies mainly on the channel bandwidth between the edge server and end devices. Cooperative transmission~\cite{gesbert2010multi} becomes promising for communication-efficient inference results delivery.
\end{itemize}

To support those computation-tasks on the resource-limited end devices, edge server inference stands out as a viable solution to fulfill the key performance requirements. The main focus of this paper is on the AI model inference for mobile devices with the edge server inference architecture.  For the edge AI inference system, energy efficiency is a key performance indicator~\cite{zhou2019edge}, which motives us to focus on the energy-efficient edge inference design. This is achieved by optimizing the overall network power consumption, including computation power consumption for performing inference tasks and transmission power consumption for returning inference results. In particular, cooperative transmission~\cite{gesbert2010multi} is a widely recognized technique to reduce the downlink transmit power consumption and provide low-latency transmission services by exploiting the high beamforming gains for edge AI inference. In this work, we thus consider that multiple edge base stations (BSs) collaboratively transmit the inference results to the  end devices~\cite{hua2019reconfigurable}. To enable transmission cooperation, we apply the computation replication principle~\cite{li2019exploiting}, i.e., the inference tasks from end devices can be performed by several neighboring edge BSs to create multiple copies of the inference results. However, computation replication greatly increases the power consumption in performing inference tasks. Therefore, it is necessary to select the inference tasks to be performed by each edge BS to achieve an optimal balance between communication and computation power consumption. 

In this paper, we propose a joint inference task selection  and downlink beamforming strategy towards achieving energy-efficient edge AI inference by optimizing the overall network power consumption consisting of the computation power consumption and the transport network power consumption under the  quality-of-service (QoS) constraints. However, the resulting formulation contains combinatorial variables and nonconvex constraints,  which makes it computationally intractable. To address this issue, we observe that the transmit beamforming vector has an intrinsic connection with the set of inference task selection (i.e., tasks are opted by edge servers to execute). Based on this crucial observation, we present a group sparse beamforming (GSBF) reformulation, followed by proposing a log-sum function based three-stage GSBF approach. In particular, in the first stage, we adopt a weighted log-sum function based relaxation to enhance the group sparsity of the structural solutions. 

Nonetheless, the log-sum function minimization problem poses challenges in computation and analysis. To resolve the issues, we present a proximal iteratively reweighted algorithm, which solves a sequence of weighted convex subproblems. Moreover, we establish the global convergence analysis and worst-case convergence rate analysis of the presented proximal iteratively reweighted algorithm. Specifically, by leveraging the Fr{\'e}chet subdifferential~\cite{kruger2003frechet}, we characterize the first-order necessary optimality conditions of the formulated convex-constrained log-sum problem. We then show that the generated iterates of the proposed algorithm make the function values steadily decrease and prove that any cluster point of the generated entire sequence is a critical point of the initial objective for any initial feasible point. Finally, we show that the defined optimality residual has $O(1/t)$ ergodic worst-case convergence rate, where $t$ is the iteration counter.

In the following, we summarize the major contributions of this paper as follows.
\begin{itemize}
        \item[1.] We propose a joint task selection  and downlink beamforming strategy to optimize the trade-off between computation and communication power consumption for an energy-efficient edge AI inference system. In particular, task selection is achieved by controlling the group sparsity structure of the transmit beamforming vector, thereby formulating a group sparse beamforming problem under the target QoS constraints.
        
        \item[2.] To solve the resulting optimization problem, we proposed a log-sum function based three-stage GSBF approach.  In particular, we adopt a weighted log-sum approximation to enhance the group sparsity of the transmit beamforming vector in the first stage. Moreover, we propose a proximal iteratively reweighted algorithm to solve the log-sum minimization problem. 
        
        \item[3.] For the presented proximal iteratively reweighted algorithm, we establish the global convergence analysis. We prove that every cluster point generated by the presented algorithm satisfies the first-order necessary
        optimality condition for the original nonconvex log-sum problem. Furthermore, a worst-case $O(1/t)$ convergence rate is established for this algorithm in an ergodic sense.

        \item[4.] Numerical experiments are conducted to demonstrate the effectiveness and competitive performance of the log-sum function based three-stage GSBF approach for designing the green edge AI inference system. 
\end{itemize}

\subsection{Related Works}                  
The study of inducing sparsity generally falls into the sparse optimization category~\cite{bach2012optimization,jenatton2011structured}. In particular, sparse optimization, emerging as a powerful tool, has recently contributed to the effective design of wireless networks, e.g., group sparse beamforming for  energy-efficient cloud radio access networks~\cite{shi2014group,dai2016energy}, and sparse signal processing for Internet-of-Things (IoT) networks~\cite{liu2018sparse,shi2020Lowoverhead}. In particular, to induce the group sparsity structure of the beamforming vector, the work of~\cite{hua2019reconfigurable, yang2019energy} adopted the mixed $\ell_{1,2}$-norm. As illustrated in~\cite{bach2012optimization}, the mixed $\ell_{1,q}$-norms ($q>1$) can induce the group sparsity structure of the interested solution. Moreover, the mixed $\ell_{1,2}$-norm and $\ell_{1,\infty}$-norm~\cite{mehanna2013joint} are commonly adopted. However, the effectiveness of sparsity based on convex sparsity-inducing norms is not satisfactory since there always exists some small nonzero elements in the obtained solutions~\cite{candes2008enhancing}. In contrast to these works, some works applied nonconvex sparsity-inducing functions to seek sparser solutions~\cite{shi2016smoothed}. Notably, the work~\cite{candes2008enhancing} reported the capability of log-sum function for enhancing the sparsity of the solutions.


 Motivated by their superior performance on inducing sparsity, we adopt log-sum functions to promote the sparsity pattern in the solutions. However, adopting the log-sum function to enhance sparsity usually makes the problem difficult to compute and analyze. In~\cite{candes2008enhancing}, the authors first proposed an iteratively reweighted $\ell_{1}$ algorithm (IRL1) for tackling the nonconvex and nonsmooth log-sum functions with linear constraints. Nonetheless, they did not further conduct the convergence analysis for the proposed method. Under reasonable assumptions, the work of~\cite{ochs2015iteratively} established the convergence results for a class of unconstrained nonconvex nonsmooth problems based on the limiting-subgradient tool. In particular, these results could apply to the log-sum model in an unconstrained setting. In~\cite{lu2014proximal}, they proposed a proximal iteratively reweighted algorithm and proved that any accumulation point is a critical point. The work of~\cite{sun2017global} further showed that, for any starting feasible points, the sequence generated by their proximal iteratively reweighted algorithm could converge to a critical point under the Kurdyka-\L{}ojasiewicz (KL) property~\cite{absil2005convergence}. However, these works focused on the unconstrained formulation or linearly constrained cases when the log-sum model is involved. The theoretical analysis for the log-sum function with general convex-set constraints has not been investigated.

\subsection{Organization}
The remainder of this paper is organized as follows. Section \ref{Sec: System-model} presents the system model of the edge AI inference, followed by the problem formulation and analysis. Section \ref{sec:3} provides the group sparse beamforming formulation. The log-sum function based three-stage GSBF approach is proposed in~ Section \ref{Sec: Log-sum based 3-stage GSBF}. Section \ref{Sec: convergence analysis} provides
the global convergence and convergence rate analysis of the proposed proximal iteratively reweighted algorithm. Section \ref{Sec: Simulation Results} demonstrates the performance of the proposed approach. The conclusion remark is made in Section \ref{Sec: conclusion}.  To keep the main text coherent and free of technical details, we divert most of the mathematic proofs to the Appendices.

\subsection{Notation}
 Throughout this paper, we subsume the notation used as follows. We use $\mathbb{C}^{n}$ and $\mathbb{R}^{n}$ to denote the complex vector space and the real Euclidean $n$-space $\mathbb{R}^{n}$, respectively. Boldface lower-case letters and upper case letters to represent vectors (e.g., $\bm{x}$) and matrices (e.g., $\bm{I}$) with an appropriate size, respectively. The inner product between $\bm{x},\bm{y}\in\mathbb{C}^{n}$ is denoted as $\langle\bm{x},\bm{y}\rangle$. $\Vert
\cdot\Vert_{1}$ and $\Vert\cdot\Vert_{2}$ is the conventionally defined $\ell_{1}$-norm and $\ell_{2}$-norm for any vectors in $\mathbb{C}^{n}$, respectively. In addition, we use $(\cdot)^{\sf{H}}$ and $(\cdot)^{\sf{T}}$ to denote the Hermitian and transpose operators, respectively. $\mathfrak{R}(\cdotp)$ is the real part of a complex scalar. $\bm{1}$ is a vector with all components equal to 1  and $\bm{0}$ denotes the zero vector with an appropriate size. In particular, $\Vert\bm{v}\Vert_{\mathcal{G}_2} = \left[\|\bm{v}_{1}\|_{2},\cdots,\|\bm{v}_{n}\|_{2}\right]^{\sf{T}} \in \mathbb{R}^{n} $ represents a vector whose $n$th element is the $\ell_{2}$-norm of a structured vector $\bm{v}_{n} \in \mathbb{C}^{n}$. We use $\circ$ to denote composition operation between two functions and symbol $\odot$ defines the elementwise product for any two vectors $\bm{x},\bm{y}\in\mathbb{C}^{n}$. 

For any closed convex set $\mathcal{C}\subset\mathbb{C}^{n}$, we use $\delta_{\mathcal{C}}(\bm{c})$ to denote the characteristic function associated with $\mathcal{C}$, which is defined as 
\begin{equation*}
\delta_{\mathcal{C}}(\bm{c}) := \left\lbrace \begin{array}{lc}
0, & \bm{c}\in\mathcal{C},\\
+\infty, & \bm{c}\notin\mathcal{C}.
\end{array}\right. 
\end{equation*}
Similarly, $\mathbb{I}(\cdot)$ defines a indicator function associated with the given condition $\cdot$, i.e., if condition $\cdot$ is met, then return the value $1$; otherwise, return the value $0$. Moreover, $z \sim\mathcal{CN}(\mu,\sigma^{2})$
corresponds to the complex random variable $z$ with mean $\mu$ and variance $\sigma^{2}$.

\section{System Model and Problem Formulation}\label{Sec: System-model}
This section describes the overall system model and power consumption model for performing intelligent tasks in the considered edge AI inference system, followed by the problem formulation and analysis.
\subsection{System Model}
\begin{figure}[!t]
	\centering
    \includegraphics[width=3.2in]{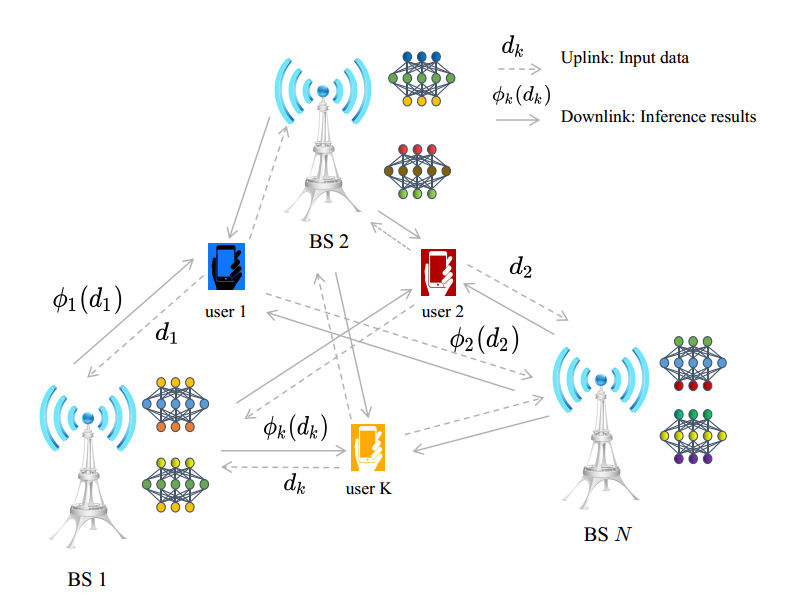}
    \caption{System model illustration of the edge AI inference for intelligent tasks. This paper considers the scenario that each neighboring edge BS has collected the raw input data $\{\bm{d}_{k}\}$ from mobile users.}
    \label{fig:sysModel}
\end{figure}

We consider an edge computing system consisting of $N$ $L$-antenna BSs collaboratively serving $K$ single-antenna mobile users (MUs), as illustrated in Fig.~\ref{fig:sysModel}. These deployed BSs are used as dedicated edge nodes and have access to the enormous computation and storage resources~\cite{mao2017survey}.  For convenience, define $\mathcal{K}=\{1,\dots,K\}$ and $\mathcal{N}=\{1,\dots,N\}$ as the index sets of MUs and BSs, respectively.
MUs have inference computing tasks, and the results can be inferred from task-related DNNs. For ease of expression, we use $\bm{d}_{k}$ to denote the raw input data collected from MU $k$, and the corresponding inference results are represented as $\phi_k(\bm{d}_k)$. As performing intelligent tasks on DNNs are typically resource-demanding, it is usually impractical to perform the tasks on resource-constrained mobile devices locally.  In the proposed edge AI inference system, by exploiting the computation replication~\cite{li2019exploiting}, we consider the scenario that each neighboring edge BS has collected the raw input data $\{\bm{d}_{k}\}$ from all MUs. Then the edge BSs process the data $\{\bm{d}_{k}\}$ for model inference. After the edge BSs complete the model inference, the inference results $\{\phi_k(\bm{d}_k)\}$ are returned to the corresponding MUs via the downlink channels. We assume that all edge BSs have been equipped with the pre-trained deep network models for all inference tasks~\cite{yang2019energy}.

In the downlink transmission, the edge BSs, which perform the inference tasks for the same MU cooperatively, return the inference results to the MU. We assume perfect channel state information (CSI) is available to all edge BSs to enable cooperative transmission for the inference results \cite{gesbert2010multi}. Let $\mathcal{A}_n\subseteq\mathcal{K}$ denote the indexes of MUs whose tasks are selectively performed by BS $n$, and  $\mathcal{A}=(\mathcal{A}_1,\cdots,\mathcal{A}_N)$ represents task selection strategy. 
\subsubsection{Downlink Transmission Model}
Let $s_{k}\in\mathbb{C}$ denote the encoded scalar of the  requested output $\{\phi_k(\bm{d}_k)\}$ for MU $k$, and $\bm{v}_{nk}\in\mathbb{C}^{L}$ be the transmit beamforming vector at the BS $n$ for $s_{k}$. For convenience, and without loss of generality, we assume that $\mathbb{E}(|s_k|^2)=1$, i.e., the power of $s_{k}$ is normalized to the unit. The transmitted signal $\bm{x}_{n}$ at BS~$n$ can be  expressed as
\begin{equation}
\bm{x}_{n} = \sum_{k\in\mathcal{A}_n}\bm{v}_{nk}s_{k}.
\end{equation}

Let $\bm{h}_{nk}\in\mathbb{C}^{L}$ be the propagation channel coefficient vector between BS $n$ and MU~$k$.
The received signal at MU~$k$ denoted as $y_k$, is then given by
\begin{equation}\label{eq: received signal model}
\begin{aligned}
y_k &= \sum_{n\in\mathcal{N}}\bm{h}_{nk}^{\sf{H}}\bm{x}_{n} + z_{k}\\
&=\sum_{n\in\mathcal{N}}\bm{h}_{nk}^{\sf{H}}\sum_{l\in \mathcal{A}_{n}}\bm{v}_{nl}s_l+z_k\\
&= \sum_{n\in\mathcal{N}} \bm{h}_{nk}^{\sf{H}} \!\left[\mathbb{I}(k\in\mathcal{A}_{n})\bm{v}_{nk}s_k \!+\! \sum_{l\in\mathcal{A}_{n},l\neq k}\!\bm{v}_{nl}s_l\right] + z_{k},
\end{aligned}
\end{equation}
where $z_k\sim\mathcal{CN}(0,\sigma^2_{k})$ is the isotropic additive white Gaussian noise.

We assume that all data symbols $s_k$ are mutually independent of each other as well as noise. Based on~\eqref{eq: received signal model}, the signal-to-interference-plus-noise ratio (SINR) for MU~$k$ is therefore given as
\begin{equation}\label{eq: downlink_sinr}
\textrm{SINR}_k(\A) = \frac{\vert\sum_{n\in\N}\mathbb{I}(k\in\A_n)\bm{h}_{nk}^{\sf{H}}\bm{v}_{nk}\vert^2}{\sum_{l\ne k}|\sum_{n\in\N}\mathbb{I}(l\in\A_n)\bm{h}_{nk}^{\sf{H}}\bm{v}_{nl}|^2+\sigma_k^2}.
\end{equation}
\subsubsection{Power Consumption Model}
The computation and transmission power consumption for model inference is generally large. Energy efficiency is of significant importance for an energy-efficient edge AI inference system design, for which the overall network power consumed in computation and communication at the edge BSs becomes our main interest. Specifically, we express the total transmission power for all edge BSs in the downlink as 
\begin{equation}\label{eq:transmission_power}
\centering
\begin{aligned}
P_{\textrm{trans}}(\mathcal{A},\{\bm{v}_{nk}\}):&=\sum_{n=1}^{N}\frac{1}{\eta_{n}}\mathbb{E}\bigg[ \sum_{k\in\mathcal{A}_{n}}\|\bm{v}_{nk}s_{k}\|_2^2\bigg]\\
&=\sum_{n=1}^{N} \sum_{k\in\mathcal{A}_{n}}\frac{1}{\eta_{n}}\|\bm{v}_{nk}\|_2^2,
\end{aligned}
\end{equation}
where $\eta_n$ is the radio frequency power amplifier efficiency coefficient of edge BS $n$.

In addition to the downlink transmission power consumption, the power consumed in performing AI inference tasks should be taken into consideration as well, owing to the power-demanding nature of running DNNs. We use $P_{nk}^{\sf{c}}$ to denote the computation power consumption of the  BS $n$ in performing inference task $\phi_k$. Then the computation power consumed by all BSs are given by
\begin{equation}
P_{\textrm{comp}}(\mathcal{A}) :=\sum_{n\in\N}\sum_{k\in\A_n}P_{nk}^{\sf{c}}.
\end{equation}

\begin{figure}[!t]
        \centering
        \includegraphics[width=3.2in]{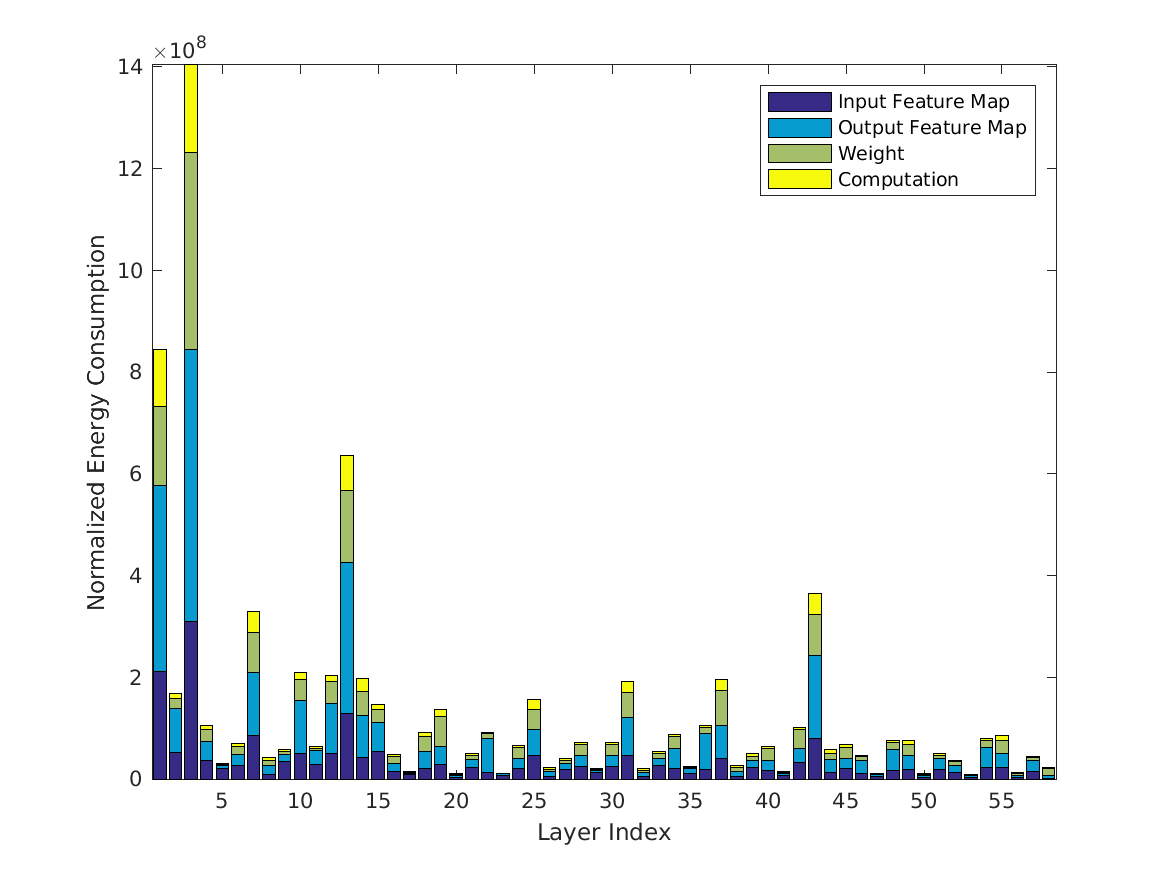}
        \caption{The estimated energy consumption breakdown~\cite{EstimationWebsite} of the GoogLeNet~v1 to perform image classification tasks on the Eyeriss chip~\cite{chen2016eyeriss}.}
        \label{fig:sysModel2}
\end{figure}

For the estimation of the computation energy consumption in executing task~$\phi_k$ therein, the works~\cite{yang2017method,yang2017designing} stated that the energy consumption of a deep neural network layer for inference mainly including computation energy consumption and data movement energy consumption. For illustration, we take GoogLeNet~v1~\cite{szegedy2015going} as a concrete example to illustrate the energy consumed by performing inference tasks. Specifically,  we use GoogLeNet~v1 to perform image classification tasks on the Eyeriss chip~\cite{chen2016eyeriss}. With the help of an energy estimation online tool~\cite{EstimationWebsite}, we are able to visualize the energy consumption breakdown of the GoogLeNet~v1, as illustrated in Fig.~\ref{fig:sysModel2}. We obtain the estimation of the computation power consumption via dividing the total energy consumption by the computation time. In particular, the computation time is determined by the total number of multiplication-and-accumulation (MAC) operations and the peak throughout of Eyeriss chip.

Therefore, the overall power consumption for edge AI inference, including transmission and computation power consumption, is calculated as
\begin{equation}\label{eq: Power_all 1}
\hspace{-0.5em}
\centering
\begin{aligned}
P_{\textrm{overall}}(\mathcal{A},\{\bm{v}_{nk}\})&\!=\! P_{\textrm{trans}}(\mathcal{A},\{\bm{v}_{nk}\}) +P_{\textrm{comp}}(\mathcal{A})\\
&=\!\sum_{n\in\mathcal{N}}\!\sum_{k\in\A_{n}}\!\frac{1}{\eta_{n}}\|\bm{v}_{nk}\|_2^2 \!+\! \!\sum_{n\in\mathcal{N}}\!\sum_{{k\in\mathcal{A}_n}} \!P_{nk}^{\sf{c}}.
\end{aligned}
\end{equation}
\subsection{Problem Formulation and Analysis}

Note that there is a fundamental trade-off between transmission and computation power consumption. To be specific, more edge BSs performing the same task for MUs can significantly reduce the transmission power by exploiting higher transmit beamforming gains. However, this inevitably increases the computation power consumption for performing inference tasks. Therefore, the goal of an energy-efficient edge inference system can be achieved by minimizing the overall network power consumption to reach a balance between these two parts of power consumption.


Let $\{\gamma_k\}$ be the target SINR for MUs to receive the reliable AI inference results in the downlink successfully. In our proposed energy-efficient edge AI inference system, the overall power minimization problem is thus formulated as
\begin{equation}\label{eq: min_sum_power}
\begin{aligned}
\min_{\substack{\mathcal{A},\{\bm{v}_{nk}\}}}~& P_{\textrm{overall}}(\mathcal{A},\{\bm{v}_{nk}\})\\ 
\textrm{s.t.}~ \ 
~\ &\text{SINR}_{k}(\mathcal{A}) \geq \gamma_k,~\forall\,k \in \mathcal{K},\\
~&\sum_{k\in\mathcal{K}}\|\bm{v}_{nk}\|_2^2\leq P_{n}^{\sf{max}},~\forall\,n\in\mathcal{N},
\end{aligned}
\end{equation}
where $P_{n}^{\sf{max}}>0$ denotes the maximum transmit power of edge BS $n$. 

Unfortunately, problem~\eqref{eq: min_sum_power} turns out to be a mixed combinatorial optimization problem due to the presence of combinatorial variable $\mathcal{A}$, which makes it computationally intractable~\cite{shen2019lorm}. On the other hand, the nonconvex SINR constraints also pose troublesome challenges for solving~\eqref{eq: min_sum_power}. To address these issues, we recast problem~\eqref{eq: min_sum_power} into a tractable formulation by inducing the group sparsity of the beamforming vector in the following section.


\subsection{A Group Sparse Beamforming Representation Framework}\label{sec:3}

One naive approach to cope with the combinatorial variable~$\mathcal{A}$ is the exhaustive search. However, it is often computationally prohibitive owing to the exponential complexity. As a practical alternative, there is a critical observation that such a combinatorial variable $\mathcal{A}$ can be eliminated by exploiting the inherent connection between task selection and the group sparsity structure of  beamforming vectors. Specifically, if edge BS $n$ does not perform the inference tasks from MU~$k$ (i.e., $k\notin \A_n$), then it will not deliver the inference result $\phi_k(\bm{d}_k)$ in the downlink transmission (i.e., $\bm{v}_{nk} = \bm{0}$). In other words, if $k\notin\A_n$, all coefficients in the beamforming vector $\bm{v}_{nk}$ are zero simultaneously. Mathematically, we have $\A_{n} = \{k\mid\bm{v}_{nk} \neq\bm{0},~k\in\K\}$, for all $n\in\N$, meaning the task selection strategy $\mathcal{A}$ can be uniquely determined by the group sparsity structure of $\bm{v}_{nk}$. In this respect, the overall network power consumption problem~\eqref{eq: min_sum_power} can rewritten as 
\begin{equation}
P_{\textrm{sparse}}(\{\bm{v}_{nk}\}) \!=\! \sum_{n=1}^{N}\!\sum_{k=1}^{K}\!\frac{1}{\eta_n}\|\bm{v}_{nk}\|_2^2\!+\!\sum_{n=1}^{N}\!\sum_{k=1}^{K} \mathbb{I}(\bm{v}_{nk} \neq\bm{0})P_{nk}^{\sf{c}}.
\end{equation}
By considering the sparsity structure in the beamforming vectors, the SINR expression~\eqref{eq: downlink_sinr} is transformed into
\begin{equation}
\begin{aligned}
\text{SINR}_k &= \frac{\vert\sum_{n\in\N}\bm{h}_{nk}^{\sf{H}}\bm{v}_{nk}\vert^2}{\sum_{l\ne k}|\sum_{n\in\N}\bm{h}_{nk}^{\sf{H}}\bm{v}_{nl}|^2+\sigma^2_{k}}\\
&=\frac{\vert\bm{h}_{k}^{\sf{H}}\bm{v}_{k}\vert^2}{\sum_{l\ne k}|\bm{h}_{k}^{\sf{H}}\bm{v}_{l}|^2+\sigma^2_{k}},~\,\forall\,k\in\K,
\end{aligned}
\end{equation}
where $\bm{h}_{k} = \left[\bm{h}_{1k}^{\sf{H}},\dots,\bm{h}_{Nk}^{\sf{H}}\right]^{\sf{H}}$ and $\bm{v}_{k}\!=\! \left[\bm{v}_{1k}^{\sf{T}},\dots,\bm{v}_{Nk}^{\sf{T}}\right]^{\sf{T}}$ are the aggregated channel vector and downlink transmit beamforming vector for MU $k$, respectively.

On the other hand, since an arbitrary phase rotation of the transmit beamforming vectors $\{\bm{v}_{nk}\}$ does not affect the downlink SINR constraints and the objective function value, we can always find proper phases to equivalently transform the SINR constraints in~\eqref{eq: min_sum_power} into convex second-order cone constraints \cite{wiesel2005linear}.
We thus have the following convex-constrained sparse optimization framework for network power minimization 
\begin{equation}\label{eq: model_total}
\begin{aligned}
\min_{\{\bm{v}_{nk}\}}~&  P_{\textrm{sparse}}(\{\bm{v}_{nk}\})\\ 
\text{s.t.} \ 
~~&\sum_{k\in\mathcal{K}}\|\bm{v}_{nk}\|_2^2\leq P_{n}^{\sf{max}},~\forall n\in\mathcal{N},\\
~&\sqrt{\sum_{l\ne k}\!|\bm{h}_k^{\sf{H}}\bm{v}_l|^2+\!\sigma^2_{k}} \!\leq \frac{1}{\sqrt{\gamma_k}}\mathfrak{R}(\bm{h}_k^{\sf{H}}\bm{v}_k),~\forall k \in \mathcal{K}.
\end{aligned}
\end{equation}
However, problem \eqref{eq: model_total} is still nonconvex due to the indicator function in the objective function. As presented in~\cite[Proposition 1]{shi2014group}, a weighted mixed $\ell_{1,2}$-norm can be served as the tightest convex surrogate of the objective in~\eqref{eq: model_total}, i.e.,
\begin{equation}
        P(\{\bm{v}_{nk}\}) = 2\sum_{n=1}^{N}\sum_{k=1}^{K}\sqrt{P_{nk}^{\sf{c}}/\eta_{n}}\|\bm{v}_{nk}\|_{2}.
\end{equation} 

In this paper, we instead propose to adopt a new group sparsity inducing function for inference tasks selection via enhancing sparsity, thereby further reducing the network power consumption.
%
\section{A Los-sum Function Based Three-stage Group Sparse Beamforming Framework}\label{Sec: Log-sum based 3-stage GSBF}
In this section, we shall propose to adopt the log-sum function to enhance the group sparsity of the beamforming vector, followed by describing the log-sum function based three-stage GSBF approach. In particular, we propose a proximal iteratively reweighted algorithm to address the log-sum minimization problem in the first stage. 
\subsection{Log-sum Function for Enhancing Group Sparsity }
Let $\bm{v} =\! \left[ \bm{v}_{11}^{\sf{T}},\cdots,\bm{v}_{1K}^{\sf{T}},\cdots,\bm{v}_{N1}^{\sf{\sf{T}}},\cdots,\bm{v}_{NK}^{\sf{T}}\right]^{\sf{T}} \!\in\!\mathbb{C}^{LNK}$ denote the aggregated beamforming vector $\bm{v}$. To promote the group sparsity for the beamforming vector
$\bm{v}_{nk}$, in this paper, we propose to use the following weighted nonconvex log-sum function as an approximation for the objective $P_{\textrm{sparse}}(\bm{v})$
\begin{equation}
\Omega(\bm{v}):= \sum_{n=1}^{N}\sum_{k=1}^{K}\rho_{nk}\log(1+p\|\bm{v}_{nk}\|_{2}),
\end{equation}
where $\rho_{nk} = \sqrt{P^{\sf{c}}_{nk}/\eta_{n}} >0$ is a weight coefficient and $p >0 $ is a tunable parameter. The main motivation for adopting such a log-sum penalty among various types of sparsity-inducing functions~\cite{bach2012optimization} is based on the following considerations:
\begin{itemize}
        \item The mixed $\ell_{1,2}$-norm is similar as an $\ell_{1}$-norm of vector $\bm{v}$ and thereof offers the tightest convex relaxation to the $\ell_0$-norm. In contrast to the mixed $\ell_{1,2}$-norm, it has been reported that the log-sum function can significantly enhance the sparsity of the solution than the conventional $\ell_{1}$-norm~\cite{bach2012optimization,candes2008enhancing}.
        
        \item From the perspective of performance and theoretical analysis of the designed algorithm, a log-sum function brings more practicability due to its coercivity and boundedness of its first derivative.
\end{itemize}
\subsection{A Log-sum Function Based Three-stage Group Sparse Beamforming Approach}
We present the proposed log-sum based three-stage GSBF framework. Specifically, the first stage is to solve the log-sum convex-constrained problem via the proposed proximal iteratively reweighted algorithm to obtain a solution $\bm{v}_{\textrm{sparse}}$; the second stage prioritizes the tasks in progress based on the obtained solution~$\bm{v}_{\textrm{sparse}}$ and  system parameters, followed by obtaining the optimal task selection strategy $\mathcal{A}^{*}$;  with fixed $\mathcal{A}^{*}$, we refine the $\bm{v}$ in the third stage. Details are depicted as follows. 

\textbf{Stage 1: Log-sum Function Minimization.}  In this first stage, we obtain the group sparsity structure of beamformer $\bm{v}$ by solving the following nonconvex program 
\begin{eqnarray}\label{eq:p_induce}
\setlength\arraycolsep{1pt}
\begin{aligned}
\min_{\bm{v}}~ \Omega(\bm{v})\quad\textrm{s.t.}\quad\bm{v}\in\mathcal{C},
\end{aligned}
\end{eqnarray}
where $\mathcal{C}$ denotes the convex-constraints in~\eqref{eq: model_total}. 

However, the nonconvex and nonsmooth objective in~\eqref{eq:p_induce} and the presence of the convex constraints usually pose challenges in computation and analysis. Inspired by the work of~\cite{candes2008enhancing}, we can iteratively minimize the objective by solving a sequence of tractable convex subproblems. The main idea of our presented algorithm is to solve a well-constructed convex surrogate subproblem instead of directly solving the original nonconvex problem. 

Let $f_{p}(\bm{v}_{nk}):= \log(1+p\Vert\bm{v}_{nk}\Vert_{2})$. First observe that $f_{p}(\bm{v}_{nk})$ is a composite function with $\bm{z}(\bm{v}_{nk})  = \|{\bm{v}}_{nk}\|_2$ convex and $f_{p}(\bm{z}) = \log(1+pz_{nk})$ nonconvex. At the $i$th iterate $\bm{v}_{nk}^{[i]}$, for any feasible $\tilde{\bm{v}}_{nk}$, we have
\begin{equation}
\begin{aligned}
f_{p}(\bm{z}(\tilde{\bm{v}}_{nk})) &\leq f_{p}(\bm{z}(\bm{v}_{nk}^{[i]})) + \langle\bm{w}^{[i]},\bm{z}(\tilde{\bm{v}}_{nk})-\bm{z}(\bm{v}_{nk}^{[i]})\rangle\\
&\leq f_{p}(\bm{z}(\bm{v}_{nk}^{[i]})) + \langle\bm{w}^{[i]},\bm{z}(\tilde{\bm{v}}_{nk})-\bm{z}(\bm{v}_{nk}^{[i]})\rangle\\ &\quad+ \frac{\beta}{2}\Vert\tilde{\bm{v}}_{nk} - \bm{v}_{nk}^{[i]}\Vert_{2}^{2},
\end{aligned}
\end{equation}
where $\bm{w}^{[i]} \in \partial(f_{p}(\bm{z}(\bm{v}_{nk}^{[i]})))$ is the subgradient of $f_{p}(\bm{z})$ at $\bm{z} = \bm{z}(\bm{v}_{nk}^{[i]})$ and $\beta > 0$ is the prescribed proximity parameter, and the first inequality holds by the definition of the subgradient of the convex function. Hence, a convex subproblem is derived as an approximation of $f_{p}(\bm{v}_{nk})$ at current iterate $\bm{v}_{nk}^{[i]}$, which reads
\begin{equation}\label{eq: convex_subproblem}
\hspace{-0.3em}
\begin{aligned}
\min_{\{\bm{v}_{nk}\}}\quad& \sum_{n=1}^{N} \sum_{k=1}^{K} w_{nk}^{[i]}\Vert\bm{v}_{nk}\Vert_{2} + \frac{\beta}{2}\sum_{n=1}^{N}\sum_{k=1}^{K}\Vert\bm{v}_{nk} -\bm{v}_{nk}^{[i]}\Vert_{2}^{2}\\
\textrm{s.t.}~\quad&\bm{v}\in\mathcal{C}
\end{aligned}
\end{equation} 
with weights
\begin{equation}\label{eq: update_weights}
w_{nk}^{[i]} = \rho_{nk}\cdot \partial(f_{p}(\bm{z}(\bm{v}_{nk}^{[i]})))         =\frac{p\rho_{nk}}{{p\Vert\bm{v}_{nk}^{[i]}\Vert_{2}} + 1}.
\end{equation}

As presented in~\cite{candes2008enhancing}, a smaller $\|\bm{v}_{nk}^{[i]}\|_2$ causes larger $w_{nk}^{[i]}$, then drive the nonzero components of $\bm{v}_{nk}$ towards zero aggressively. Overall, to enhance the group sparsity structure of the beamforming vector, the proposed proximal iteratively reweighted algorithm is illustrated in  Algorithm~\ref{alg.0}.
\begin{algorithm}
        \caption{\small A Proximal Iteratively Reweighted Algorithm for Solving~\eqref{eq:p_induce}}
        \label{alg.0}
        \begin{algorithmic}[1]
                \STATE \textbf{Input: } $p > 0$, $\beta > 0$, $\textrm{IterMax}$ and $\bm{v}^{[0]}$.
                \STATE \textbf{Initialization:} $\bm{w}^{[0]} = \bm{1}$ and set $i = 0$.
                \WHILE{\textrm{not converge or not attain IterMax}}
                \STATE (Solve the reweighted subproblem for $\bm{v}^{[i+1]}$) Calculate $\bm{v}^{[i+1]}$ according to~\eqref{eq: convex_subproblem}.
                \STATE {(Reweighting) Update weight $w_{nk}^{[i]}$ according to~\eqref{eq: update_weights}.}
                \STATE {Set $i \leftarrow i+1$.}
                \ENDWHILE
                \STATE \textbf{Output: } $\bm{v}_{\textrm{sparse}} = \bm{v}^{[i]}$.
        \end{algorithmic}
\end{algorithm}

\textbf{Stage 2: Tasks Selection.}
In this second stage, an ordering guideline is applied to determine the priority of inference tasks, which is guided by the solution obtained in~\textbf{Stage 1}. For ease of notation, let $\mathcal{S}=\{(n,k)\mid n\in\mathcal{N}, k\in\mathcal{K}\}$ denote the set of all tasks. By considering the key system parameters (e.g., $\bm{h}_{nk}$, $P_{nk}^{\sf{c}}$ and $\eta_{n}$), the priority of task $(n,k)$ is heuristically given as 
\begin{equation}\label{eq: guide_line}
\theta_{nk} = \sqrt{\frac{\|\bm{h}_{nk}\|_2^2\eta_{n}}{{P}_{nk}^{\sf{c}}}}\|\bm{v}_{nk} \|_2.
\end{equation}

Intuitively, if edge BS $n$ is with a lower aggregative beamformer gain, lower power amplifier efficiency, lower channel power gain, but a higher computation power consumption for MU~$k$, task $(n,k)$  has a lower priority. A lower $\theta_{nk}$ indicates that the tasks from MU $k$ have lower priority and may not be performed by BS $n$. Thus, tasks are arranged in light of the rule~\eqref{eq: guide_line}  with descending order. That is, the task's priority is $\theta_{\pi_1}\geq\theta_{\pi_2}\dots\geq\theta_{\pi_{NK}}$, where $\pi$ denotes the permutation of task indexes.

We then solve a sequence of convex feasibility detection problems to obtain task selection strategy $\mathcal{A}$,
\begin{equation}\label{prob:S2}
\begin{aligned}
\mathop{\textrm{find}}~\bm{v}\quad\textrm{s.t.}\quad\bm{v}_{\pi^{(t)}}=\bm{0},~\bm{v} \in \mathcal{C},
\end{aligned}
\end{equation}
where $\pi^{(t)}=\{\pi_{t+1},\cdots,\pi_{NK}\}$ and $t$ increases from $K$ to $NK$ until~\eqref{prob:S2} is feasible. Here $\bm{v}_{\pi^{(t)}}=\bm{0}$ are convex constraints, meaning that all $\bm{v}_{nk}$'s coefficients are zeros for task $(n,k)\in\pi^{(t)}$. The support set of beamformer $\bm{v}$ is defined as $\mathcal{T}(\bm{v}) = \{(n,k)\mid\Vert\bm{v}_{nk}\Vert_{2} \neq 0, n\in\mathcal{N},k\in\mathcal{K}\}$, then the optimal task selection strategy $\mathcal{A}^*$ can be derived from  $\mathcal{T}(\bm{v})=\{\pi_{1},\dots, \pi_{t}\}$.

\textbf{Stage 3: Solution Refinement.}
At this point, we have determined tasks selection for each BS. Then, fix the obtained task index set, we solve the following convex program to refine the beamforming vectors
\begin{equation}\label{prob:S3}
\setlength{\abovedisplayskip}{3pt}
\setlength{\belowdisplayskip}{3pt}
\begin{aligned}
\min_{\bm{v}}\quad& \sum_{n=1}^{N}\sum_{k=1}^{K}\frac{1}{\eta_{n}}\|\bm{v}_{nk}\|_2^2+\sum_{n=1}^{N}\sum_{k\in\mathcal{A}^{*}}P_{nk}^{\sf{c}}\\ 
\textrm{s.t.}\quad&\bm{v}_{\pi^{(t)}}=\bm{0},\\
&\bm{v}\in\mathcal{C}.
\end{aligned}
\end{equation}

Overall, our proposed log-sum function based three-stage GSBF framework for solving~\eqref{eq: min_sum_power} can be presented in Algorithm~\ref{alg.1}. 
\begin{algorithm}
        \caption{\small A Log-sum Function Based Three-stage GSBF Framework for Solving~\eqref{eq: min_sum_power}}
        \label{alg.1}
        \begin{algorithmic}[1]
                \STATE \textbf{Stage 1:} Log-sum function minimization. Call Algorithm~\ref{alg.0} to obtain $\bm{v}_{\textrm{sparse}}$.
                \STATE \textbf{Stage 2:}  Tasks selection.
                \STATE Sort $\theta_{nk}$'s in descending order by~\eqref{eq: guide_line}.
                \STATE Set $t=K$.
                \WHILE{\textrm{not feasible}} 
                \STATE (Determine the optimal task selection) Calculate~\eqref{prob:S2} with $\pi^{(t)} = \{\pi_{t+1}, \dots, \pi_{NK}\}$.
                \STATE {Set $t \leftarrow t+1$.}
                \ENDWHILE
                \STATE \textbf{Stage 3:}  Solution refinement.
                \STATE Refine $\{\bm{v}_{nk}\}$ by solving~\eqref{prob:S3}.
                \STATE Obtain $\{\bm{v}_{nk}\}$ and $\mathcal{A}^{*}$.
        \end{algorithmic}
\end{algorithm}

\section{Global Convergence Analysis}\label{Sec: convergence analysis}
In this section, we provide the global convergence for Algorithm~\ref{alg.0}. Specifically, we derive the first-order necessary optimality condition to characterize the optimal solutions. We then establish convergence results for a subsequence of the sequence generated by Algorithm~\ref{alg.0}. Furthermore, we show that for any initial feasible point, the entire sequence must have cluster points, and any cluster point satisfies the established first-order optimality condition. Finally, the ergodic worst-case convergence rate of the optimality residual is derived.
\subsection{First-order Necessary Optimality Condition}
In this subsection, we derive the first-order necessary conditions to characterize the optimal solution of~\eqref{eq:p_induce}. Problem~\eqref{eq:p_induce} is equivalently rewritten as
\begin{equation}\label{eq:unconstrainted}
\min_{\bm{v}}~ J(\bm{v}) := \Omega(\bm{v}) + \delta_{\mathcal{C}}(\bm{v}).
\end{equation}
Similarly, for the derived subproblem~\eqref{eq: convex_subproblem}, we have
\begin{equation}\label{eq: unconstrained_sub}
\begin{aligned}
\min_{\bm{v}}~\!G(\bm{v};\bm{v}^{[i]}):= \!\sum_{n=1}^{N}\!\sum_{k=1}^{K}\!w_{nk}^{[i]}\Vert\bm{v}_{nk}\Vert_{2} + \frac{\beta}{2}\Vert\bm{v} - \bm{v}^{[i]}\Vert_{2}^{2}\!+\! \delta_{\mathcal{C}}(\bm{v}).
\end{aligned}
\end{equation}

Due to the nonconvex and nonsmooth nature of the log-sum function, we make use of the Fr\'echet subdifferential as the major tool in our analysis. Its definition is introduced as follows.
\begin{definition}[Fr\'echet subdifferential~\cite{kruger2003frechet}]\label{Frechet subdifferentials} 
        Let $\mathcal{X}$ be a real Banach space and $\mathcal{X}^{*}$ denotes the corresponding topological dual and $f$ be a function from $\mathcal{X}$ into an extended real line $\bar{\mathbb{R}} = \mathbb{R}\cup\{+\infty\}$, finite at $\bm{r}$. A set
        \begin{equation*}
        \partial_{F}f(\bm r) \!=\! \left\lbrace \bm{r}^{*}\!\in\! \mathcal{X}^{*}|\liminf_{\bm{u}\to\bm{r}}\frac{f(\bm u)-f(\bm{r})\!-\!\langle\bm{r}^{*},\bm{u}-\bm{r}\rangle}{\|\bm{u}\!-\!\bm{r}\|_{2}}\geq 0\right\rbrace
        \end{equation*}
        is called a Fr\'echet subdifferential of $f$ at $\bm{r}$. Its elements are referred to as Fr\'echet subgradients. 
\end{definition}

Several important properties of the Fr\'echet subdifferential~\cite{kruger2003frechet} are listed below, which are used to characterize the optimal solution of~\eqref{eq:p_induce}.
\begin{proposition}\label{Prop: Frechet}
        Let $X$ be a closed and convex set. Then the following properties on Fr\'echet subdifferentials holds true. 
        \begin{itemize}
                \item[$(\RNum{1})$] If $f$ is Fr\'echet subdifferentiable at $\bm{x}$ and attains local minimum at $\bm{x}$, then
                $\bm{0} \in \partial_{F}f(\bm{x})$.
                
                \item[$(\RNum{2})$] Let $h(\cdot)$ be Fr\'echet subdifferentiable at $\bar{z} = z(\bar{\bm{x}})$ with $z(\bm{x})$ being convex, then $h \circ z(\bm{x})$ is Fr\'echet subdifferentiable at $\bar{\bm{x}}$ such that 
                \begin{equation*}
                \bar{y}\partial z(\bar{\bm{x}}) \subset \partial_{F} h\circ z(\bar{\bm{x}})
                \end{equation*}
                for any $\bar{y} \in \partial_{F}h(\bar{z})$.
                
                \item[$(\RNum{3})$]  $\mathcal{N}_{X}(\bm{x}) = \partial_{F}\delta_{X}(\bm{x})$ with closed and convex sets $X$.
        \end{itemize}
\end{proposition}  

The following Fermat's rule~\cite{rockafellar2015convex} describes the necessary optimality condition of problem~\eqref{eq:p_induce}. 
\begin{theorem}[Fermat's rule]\label{Theorem:Fermat}
        If~\eqref{eq:unconstrainted} attains a local minimum at $\bm{v}$, then it holds true that
        \begin{equation}\label{eq: Optimality_condition}
        \bm{0} \in \partial_{F}J(\bm{v}) := \partial_{F}\Omega(\bm{v}) + \mathcal{N}_{\mathcal{C}}(\bm{v}).
        \end{equation}
\end{theorem}

        We next investigate the properties of $\partial_{F}\Omega(\bm{v})$ in the following Proposition~\ref{Pro: F_derivate}, indicating that the Fr\'echet subdifferentials of $f_{p}(\cdot)$ at $\bm{0}$ is bounded.
\begin{proposition}\label{Pro: F_derivate}
        If $\bm{v}_{nk}\neq\bm{0}$, then $\gamma\partial \bm{z}(\bm{v}_{nk}) \subset\partial_{F}f_{p}\circ \bm{z}(\bm{v}_{nk})$ for any $\gamma \in \partial_{F} f_{p}(\bm{z})$. In particular, $\partial_{F} f_{p} \circ \bm{z}(\bm{0})$ is any element of $\{\bm{y}\in\mathbb{C}^{L} \mid \Vert \bm{y}\Vert_{2} \leq p\}$.
\end{proposition}

To explore the behavior of the proposed proximal iteratively reweighted algorithm, based on Theorem~\ref{Theorem:Fermat} and Proposition~\ref{Pro: F_derivate}, we define the optimality residual associated with~\eqref{eq:unconstrainted} at a point $\bm{v}^{[i]}$ as 
\begin{equation}\label{eq:Optimality_Residual}
\bm{r}^{[i]} := \bm{w}^{[i]} \odot \bm{x}^{[i]} +\bm{u}^{[i]},
\end{equation}
where $\bm{x}^{[i]} \in \partial\Vert\bm{v}^{[i]}\Vert_{\mathcal{G}_2} = \left[(\partial\|\bm{v}_{11}^{[i]}\|_{2})^{\sf{T}},\cdots,(\partial\|\bm{v}_{NK}^{[i]}\|_{2})^{\sf{T}} \right]^{\sf{T}} $ and $\bm{u}^{[i]} \in \mathcal{N}_{\mathcal{C}}(\bm{v}^{[i]})$. Since $\bm{r}^{[i]}\in \partial_{F}J(\bm{v}^{[i]})$, it implies that if $\bm{r}^{[i]} = \bm{0}$ then $\bm{v}^{[i]}$ satisfies the first-order necessary optimality condition~\eqref{eq: Optimality_condition}. We adopt $\bm{r}^{[i]}$ to measure the convergence rate of our algorithm.

Moreover, we provide the first-order optimality condition of the subproblem~\eqref{eq: unconstrained_sub} as follows
\begin{equation}\label{eq: sub-optimality}
\begin{aligned}
\bm{0} = \partial G(\bm{v};\bm{v}^{[i]}) = \beta(\bm{v}^{[i+1]} - \bm{v}^{[i]}) + \bm{w}^{[i]} \odot \bm{x}^{[i+1]} +\bm{u}^{[i+1]},
\end{aligned}
\end{equation}
where $\bm{v}^{[i+1]} = \argmin_{\bm{v}} G(\bm{v};\bm{v}^{[i]})$, $\bm{x}^{[i+1]} \in \partial\Vert\bm{v}^{[i+1]}\Vert_{\mathcal{G}_2}$ and $\bm{u}^{[i+1]} \in \mathcal{N}_{\mathcal{C}}(\bm{v}^{[i+1]})$. Note that the existence of optimal solution to~\eqref{eq: unconstrained_sub} simply follows from the convexity and the coercivity of the objective $G(\cdot;\bm{v}^{[i]})$. 

Now we show that an optimal solution of~\eqref{eq: unconstrained_sub} also satisfies the first-order necessary optimality condition of~\eqref{eq:unconstrainted} in the following lemma.

\begin{lemma}\label{lem: Lemma2}
        $\bm{v}^{[i]}$ satisfies the first-order necessary optimality condition of~\eqref{eq:unconstrainted} if and only if 
        \begin{equation*}
        \bm{v}^{[i]} = \argmin_{\bm{v}} G(\bm{v};\bm{v}^{[i]}).
        \end{equation*}
\end{lemma}
\begin{IEEEproof}
        Please refer to Appendix~\ref{app_A} for details. 
\end{IEEEproof}

Define the model reduction caused by $\bm{v}^{[i+1]}$ at a point $\bm{v}^{[i]}$ as
\begin{equation}\label{eq: model_reduction}
\Delta G(\bm{v}^{[i+1]};\bm{v}^{[i]}) := G(\bm{v}^{[i]};\bm{v}^{[i]}) - G(\bm{v}^{[i+1]};\bm{v}^{[i]}).
\end{equation}

The new iterate $\bm{v}^{[i+1]}$ causes a decrease in the objective $J(\bm{v})$, and this model reduction~\eqref{eq: model_reduction} converges to zero in the limit, both results are revealed in the following Lemma~\ref{lem: lemma3}.
\begin{lemma}\label{lem: lemma3}
        Suppose $\{\bm{v}^{[i]}\}_{i=0}^{\infty}$ is generated by of Algorithm~\ref{alg.0} with $\bm{v}^{[0]} \in \mathcal{C}$. The following statements hold true 
        \begin{itemize}
                \item[$(\RNum{1})$] $J(\bm{v}^{[i+1]}) - J(\bm{v}^{[i]}) \leq G(\bm{v}^{[i+1]};\bm{v}^{[i]}) - G(\bm{v}^{[i]};\bm{v}^{[i]}) \leq 0$.
                
                \item[$(\RNum{2})$] $\lim\limits_{i\to \infty} G(\bm{v}^{[i+1]};\bm{v}^{[i]}) = 0$.
                
                \item[$(\RNum{3})$] $G(\bm{v};\bm{v}^{[i]})$ is monotonically decreasing. Indeed,
                $$\Delta G(\bm{v}^{[i+1]};\bm{v}^{[i]}) \geq \frac{\beta}{2}\Vert\bm{v}^{[i]} - \bm{v}^{[i+1]}\Vert_{2}^{2}.$$
        \end{itemize}
\end{lemma}
\begin{IEEEproof}
        Please refer to Appendix~\ref{app_B} for details. 
\end{IEEEproof}

We now provide our main result in the following Theorem~\ref{theorem: main theorem1}.
\begin{theorem}\label{theorem: main theorem1}
        Suppose $\{\bm{v}^{[i]}\}_{i=0}^{\infty}$ is generated by Algorithm~\ref{alg.0} with $\bm{v}^{[0]} \in \mathcal{C}$. It holds true that $\{\bm{v}^{[i]}\}$ must be bounded and any cluster point of $\{\bm{v}^{[i]}\}$ satisfies the first-order necessary optimality condition of~\eqref{eq:unconstrainted}.
\end{theorem}
\begin{IEEEproof}
        Please refer to Appendix~\ref{app_C} for details. 
\end{IEEEproof}
\subsection{Ergodic Worst-case Convergence Rate}
In this subsection, we show that the presented proximal iteratively reweighted algorithm has $O(1/t)$ ergodic worst-case convergence rate in terms of the optimality residual. In the following Lemma, it states that the optimality residual has an upper bound with the displacement of the iterates.  
\begin{lemma}\label{lem: Optimality residual is bounded}
        The optimality residual associated with problem~\eqref{eq:unconstrainted} satisfies 
        \begin{equation*}\label{eq: KKT residual_upperbound_x^{k}-x^{k+1}}
        \Vert\bm{r}^{[i+1]}\Vert_{2}^{2} \leq (\beta^2 +2\beta \kappa p^2 +\kappa^{2}p^4)\|\bm{v}^{[i]} - \bm{v}^{[i+1]}\|_{2}^{2}
        \end{equation*}
        with $\kappa = \rho^{\max}$\footnote{$\rho^{\max}$ denotes the maximum elements among $\left[ \rho_{nk}\right] $ for all $n\in\N$, $k\in\K$.}.
\end{lemma}
\begin{IEEEproof}
        Please refer to Appendix~\ref{app_D} for details. 
\end{IEEEproof}

The subproblem~\eqref{eq: unconstrained_sub} is referred to as the primal problem, and by exploiting the conjugate function~\cite{rockafellar2015convex}, the associated Fenchel-Rockafellar dual is constructed as 
\begin{equation}\label{eq: sub-dual1}
\begin{aligned}
\max_{\bm{\lambda},\bm{\mu}} \quad\ & Q(\bm{\lambda},\bm{\mu};\bm{v}^{[i]})\\
\text{s.t.} \quad\ & \Vert \bm{\lambda}_{nk} \Vert_{2} \leq 1, \quad\forall n\in \mathcal{N}, k \in \mathcal{K},
\end{aligned} 
\end{equation}
where the dual objective is given as $Q(\bm{\lambda},\bm{\mu};\bm{v}^{[i]}) = -\frac{1}{2\beta}\|\bm{\lambda}\odot\bm{w}^{[i]}+\bm{\mu}-\beta\bm{v}^{[i]}\|_{2}^{2} +\frac{\beta}{2}\|\bm{v}^{[i]}\|_{2}^{2}-\delta_{\mathcal{C}}^{*}(\bm{\mu})$, and the technical details to construct~\eqref{eq: sub-dual1} is provided in Appendix~\ref{app_A0}.

The Fenchel-Rockafellar duality theorem~\cite{rockafellar2015convex} states that the solution to~\eqref{eq: sub-dual1} provides a lower bound on the minimum value to the solution of~\eqref{eq: unconstrained_sub}. Moreover, the gap between the primal objective function value of~\eqref{eq: unconstrained_sub} and the corresponding dual objective function value of~\eqref{eq: sub-dual1} at the $i$th iterate is defined as
\begin{equation}\label{eq: duality gap}
g(\bm{v},\bm{\lambda},\bm{\mu};\bm{v}^{[i]}) := G(\bm{v};\bm{v}^{[i]}) - Q(\bm{\lambda},\bm{\mu};\bm{v}^{[i]}).
\end{equation}
If this gap $g(\bm{v},\bm{\lambda},\bm{\mu};\bm{v}^{[i]})$ is zero, then the strong duality holds. That is, at the optimal solution $(\bm{v}^{[i+1]}; \bm{\lambda}^{[i+1]}, \bm{\mu}^{[i+1]})$, we have
\begin{equation}\label{eq: strong-duality}
\Delta G(\bm{v}^{[i+1]};\bm{v}^{[i]}) = g(\bm{v}^{[i+1]}, \bm{\lambda}^{[i+1]},\bm{\mu}^{[i+1]};\bm{v}^{[i]}).
\end{equation}

We now show that the duality gap vanish asymptotically in the following Theorem.
\begin{theorem}\label{theo: duality gap gone}
        Let $\{\bm{v}^{[i]}\}_{i=0}^{\infty}$ be the sequence generated by Algorithm~\ref{alg.0} with $\bm{v}^{[0]} \in \mathcal{C}$. Then $\Vert r^{[i+1]}\Vert_{2}^{2}$ has $O(\frac{1}{t})$ ergodic worst-case convergence rate.
\end{theorem}
\begin{IEEEproof}
        Please refer to Appendix~\ref{app_E} for details. 
\end{IEEEproof}
\section{Numerical Experiments}\label{Sec: Simulation Results}
In this section, we use numerical experiments to validate the effectiveness of our proposed algorithms and illustrate
the presented theoretical results. We compare the log-sum function based three-stage GSBF approach with the coordinated beamforming approach (CB)~\cite{dahrouj2010coordinated} and mixed $\ell_{1,2}$ GSBF~\cite{shi2014group} beamforming approach (Mixed $\ell_{1,2}$ GSBF). These two approaches are listed below:
\begin{itemize}
        \item \textbf{CB} considers minimizing the total transmit power consumption. In other words, all BSs are required to perform the inference tasks from all MUs. 
        
        \item \textbf{Mixed $\bm{\ell_{1,2}}$ GSBF} considers adopting the mixed $\ell_{1,2}$-norm (i.e., the objective function in~\eqref{eq:p_induce} is replaced with  $\sum_{n=1}^{N}\sum_{k=1}^{K}\rho_{nk}\|\bm{v}_{nk}\|_{2}$) to induce group sparsity of the beamforming vector in $\textbf{Stage 1}$ of Algorithm~\ref{alg.1}.
\end{itemize}

On the experimental set-up, we consider the edge AI inference system with $8$ $2$-antennas, and $15$ single-antenna MUs that all are uniformly and independently distributed in a $\left[-0.5,0.5\right]$km $\times$ $\left[-0.5,0.5\right]$km square region. The channel between BS $n$ and MU $k$ is set as $\bm{h}_{nk} = 10^{-L(d_{nk})/20}\bm{\xi}_{nk}$, where the path-loss model is given by $L(d_{nk})=128.1 + 37.6\log_{10}d_{nk}$ and $d_{nk}$ is the Euclidean distance between BS $n$ and MU $k$, $\bm{\xi}_{nk}$ is the small-scale fading coefficient, i.e., $\bm{\xi}_{nk}\sim\mathcal{CN}(\bm{0},\bm{I})$. We set $P^{\sf{c}}_{nk}=0.45$W and specify $P_n^{\max}=1$W, $\eta_n=25\%$ and $\sigma_k^2=1$. Furthermore, for the proposed log-sum function based three-stage GSBF approach, we set $p = 100$, $\beta = 0.1$ and initialize $\bm{w}^{[0]} = \bm{1}$. In particular, we terminate the proximal iteratively reweighted algorithm either it hits the predefined maximum iterations $\textrm{IterMax} = 25$ or satisfies
\begin{equation}
\|\bm{w}^{[i+1]} - \bm{w}^{[i]}\|_{1}  \leq \epsilon,
\end{equation}
where $\epsilon = 10^{-5}$ is a predescribed tolerance.
\subsection{Convergence of the Proximal Iteratively Reweighted Algorithm}
\begin{figure}[!t]
        \centering
        \includegraphics[width=3.2in]{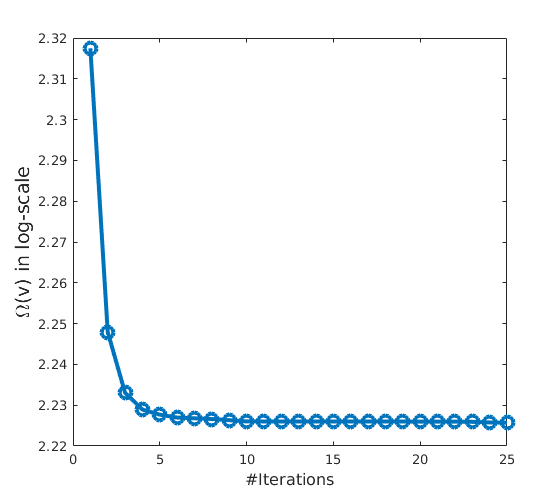}
        \caption{Convergence of the proximal iteratively reweighted algorithm for log-sum minimization problem.}
        \label{fig:convergence}
\end{figure}

The goal in this subsection is to illustrate the convergence behavior of the proposed proximal iteratively reweighted algorithm. The presented result is obtained in a typical channel realization. Fig.~\ref{fig:convergence} illustrates the convergence of the proximal iteratively reweighted algorithm. We can see that $\Omega(\bm{v})$ steadily decreases along with the iterations, which is consistent with our analysis in Lemma~\ref{lem: lemma3}. Interestingly, we observe that the objective value  of $\Omega(\bm{v})$ drops quickly in the first few iterations (less $5$ iterations), which indicates that the proposed proximal iteratively reweighted algorithm converges very fast. In view of this, we may suggest early terminating the Algorithm~\ref{alg.0} in practice to obtain an approximate solution to speed up the entire algorithm while guaranteeing the overall performance.
\subsection{Effectiveness of the Proposed Approach}
We evaluate the performance of the three algorithms in terms of the overall network power consumption, the transmit power consumption and the number of computation tasks. The presented results are averaged over $100$ randomly and independently generated channel realizations.

\begin{figure}[h!]
        \centering
        \includegraphics[width=3.2in]{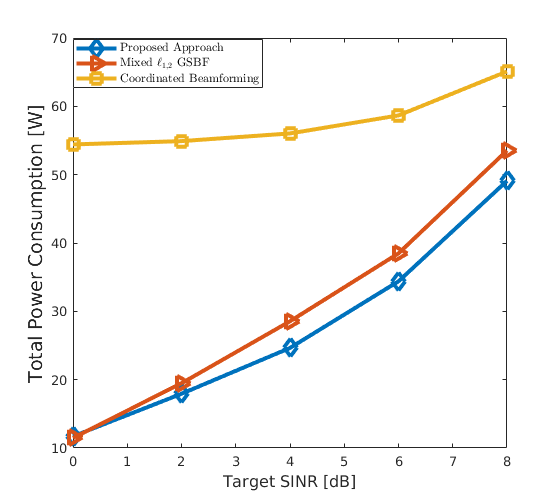}
        \caption{Average total network power consumption comparison for three different approaches in edge AI inference system.}
        \label{fig:total_power}
\end{figure}

Fig.~\ref{fig:total_power} depicts the overall network power consumption of three approaches with different target SINRs. First, we observe that all three approaches have higher total power consumption as the required SINR becomes more stringent. This is because more edge BSs are required to transmit the inference results for higher QoS. In addition, we can see that CB approach has the highest power consumption among three approaches and the relative power difference between CB and the other two approaches can achieve approximately $67\%$ when SINR is $2$dB and approximately $18\%$ when SINR is $8$dB,  indicating the effectiveness of joint task selection strategy and group sparse beamforming approach to minimize the overall network power consumption. On the other hand, we can see that the proposed log-sum function based three-stage GSBF approach outperforms the mixed $\ell_{1,2}$ GSBF approach, which demonstrates that enhance the group sparsity further reduces the overall network power consumption. In particular, we also observe that the performance gap between the blue and the red curve approximately remains at $9\%$ when SINR ranges from $4$dB to $8$dB, which indicates that the proposed log-sum function based three-stage GSBF approach is still attractive in the high SINR regime.

Tables~\ref{tab:numberoftasks} and~\ref{tab:transmission} further demonstrate the number of inference tasks performed by edge BSs and the transmission power consumption, respectively. To be specific, in Table~\ref{tab:numberoftasks}, we observe that the number of performed inference tasks among three approaches is different under various SINRs, which shows the existence of the task selection strategy. Besides, it is observed that the log-sum function based three-stage GSBF approach always achieves a less number of performed inference tasks compared to the mixed $\ell_{1,2}$ GSBF approach for target SINRs, which indicates that the log-sum function based three-stage GSBF approach can enhance the group sparsity pattern in the beamforming vector.  Meanwhile, as observed in Table~\ref{tab:transmission}, the CB approach has the lowest transmission power compared to the other two approaches because the CB approach only optimizes the power consumption in transmission with performing all inference tasks. On the other hand, the transmission power consumption of the log-sum function based three-stage GSBF approach is slightly higher compared to the mixed $\ell_{1,2}$ GSBF approach under most  SINRs. This is because more edge BSs participate in performing inference tasks in the mixed $\ell_{1,2}$ GSBF approach, resulting in a higher transmit beamforming gain for reducing transmission power. In other words, less number of performed inference tasks further reduces the computation power consumption of edge BSs but increases the transmission power consumption. Observe the Fig.~\ref{fig:total_power} and Tables~\ref{tab:numberoftasks}-\ref{tab:transmission} together, it indicates that the proposed joint task selection strategy and GSBF approach find a good balance between computation power consumption and transmission power consumption, yielding lowest network power consumption.
\begin{table}[h]
        \centering
        \caption{The Number of Performed Inference Tasks with Different Approaches}
        \label{tab:numberoftasks}
        \begin{tabular}{c| c| c| c}
                \hline
                
                \multicolumn{1}{ c| }{Target SINR~[dB]}&\multicolumn{1}{ c| }{Proposed}&\multicolumn{1}{ c| }{Mixed $\ell_{1,2}$ GSBF }&\multicolumn{1}{ c }{CB}\\
                \hline
                
                $ 0 $&$ 16.49 $&$ 17.79 $&$ 120.00 $\\
                \hline
                
                $ 2 $&$ 21.00 $&$ 22.50 $&$ 120.00 $\\
                \hline
                
                $ 4 $&$ 30.08 $&$ 34.47 $&$ 120.00 $\\
                \hline
                $ 6 $&$ 43.57 $&$ 52.17 $&$ 120.00 $\\
                \hline
                $ 8 $&$ 67.76 $&$ 77.36 $&$ 120.00 $\\
                \hline
        \end{tabular}
\end{table}
\begin{table}[h]
        \centering
        \caption{The Total Transmit Power Consumption with Different Approaches}
        \label{tab:transmission}
        \begin{tabular}{c| c| c| c}
                \hline
                \multicolumn{1}{ c| }{Target SINR~[dB]}&\multicolumn{1}{ c| }{Proposed}&\multicolumn{1}{ c| }{Mixed $\ell_{1,2}$ GSBF }&\multicolumn{1}{ c }{CB}\\
                \hline
                $ 0 $&$ 4.16 $&$ 4.68 $&$ 0.41 $\\
                \hline
                $ 2 $&$ 9.12 $&$ 8.31 $&$ 0.88 $\\
                \hline
                $ 4 $&$ 12.24 $&$ 11.76 $&$ 1.99 $\\
                \hline
                $ 6 $&$ 15.81 $&$ 14.70 $&$ 4.57 $\\
                \hline
                $ 8 $&$ 18.81 $&$ 18.58 $&$ 10.79 $\\
                \hline
        \end{tabular}
\end{table}


\section{Conclusion}\label{Sec: conclusion}
In this paper, we developed an energy-efficient edge AI inference system through the joint selection of the inference tasks and optimization of the transmit beamforming vectors for minimizing the computation power consumption and the downlink transmission power consumption, respectively. Based on the critical insight that the inference tasks selection can be achieved by controlling the group sparsity structure of  transmit beamforming vectors, we developed a group sparse optimization framework for network power minimization, for which  a log-sum function based three-stage group sparse beamforming algorithm was developed to enhance group sparsity in the solutions. To resolve the resulting nonconvex and nonsmooth log-sum function minimization problem, we further proposed a proximal iteratively reweighted algorithm. Furthermore, the global convergence analysis was provided, and a worst-case $O(1/t)$ convergence rate in an ergodic sense has been derived for this algorithm.  

\appendices
\section{Proof of Lemma~\ref{lem: Lemma2}}\label{app_A}
        Let $\bm{x}^{[i]} \in \partial\Vert\bm{v}\Vert_{\mathcal{G}_2}$ and $\bm{u}^{[i]} \in \mathcal{N}_{\mathcal{C}}(\bm{v}^{[i]})$. If $\bm{v}^{[i]} = \argmin_{\bm{v}} G(\bm{v};\bm{v}^{[i]})$, by~\eqref{eq: sub-optimality}, we have
        \begin{eqnarray}\label{eq: proof_lem2}
        \bm{0} = \bm{w}^{[i]} \odot \bm{x}^{[i]}+\bm{u}^{[i]}.
        \end{eqnarray}
        We conclude that $\bm{v}^{[i]}$ satisfies~\eqref{eq: Optimality_condition}, indicating that $\bm{v}^{[i]}$ is first-order optimal for~\eqref{eq:unconstrainted}. 
        
        Conversely, if $\bm{v}^{[i]}$ satisfies~\eqref{eq: Optimality_condition}, implying $\bm{v}^{[i]}$ satisfies~\eqref{eq: proof_lem2} by Proposition~\ref{Pro: F_derivate}. Thus $\bm{v}^{[i]}$ must be the optimal solution to the subproblem~\eqref{eq: convex_subproblem}. This completes the proof.

\section{Proof of Lemma~\ref{lem: lemma3}}\label{app_B}
        First of all, $\bm{v}^{[i+1]} = \argmin_{\bm{v}} G(\bm{v};\bm{v}^{[i]})$ and $G(\bm{v};\bm{v}^{[i]})$ is convex, so that $G(\bm{v}^{[i+1]};\bm{v}^{[i]})- G(\bm{v}^{[i]};\bm{v}^{[i]}) \leq 0$. Since  $f_{p}(\cdot)$ is concave, we have,
        \begin{eqnarray}\label{eq: f_p is concave}
        f_{p}(\bm{z}) \leq f_{p}(\bm{z}_{0}) + \langle\nabla f_{p}(\bm{z}_0),(\bm{z}-\bm{z}_0)\rangle,~\forall \bm{z},\bm{z}_0\in\mathbb{R}_{+}^{L}.
        \end{eqnarray}
        Therefore
        \begin{eqnarray}\label{eq: sub_upperbound_obj}
        \begin{aligned}
        J(\bm{v}^{[i+1]}) &= \Omega(\bm{v}^{[i+1]})\\
        &\leq \Omega(\bm{v}^{[i]}) + \sum_{n=1}^{N}\sum_{k=1}^{K} w_{nk}^{[i]}(\|\bm{v}_{nk}^{[i+1]}\|_2 -\|\bm{v}_{nk}^{[i]}\|_2 )\\
        &\quad +\frac{\beta}{2}\Vert\bm{v}^{[i+1]}-\bm{v}^{[i]}\Vert_{2}\\
        &= J(\bm{v}^{[i]}) + G(\bm{v}^{[i+1]};\bm{v}^{[i]}) - G(\bm{v}^{[i]};\bm{v}^{[i]}),
        \end{aligned}
        \end{eqnarray}
        where the first inequality follows from~\eqref{eq: f_p is concave}. This completes the first statement~$(\RNum{1})$.
        
        On the other hand, by~\eqref{eq: sub_upperbound_obj}, we have 
        \begin{eqnarray}\label{eq: sub_primal}
        \Delta G(\bm{v}^{[i+1]};\bm{v}^{[i]}) \leq J(\bm{v}^{[i]}) - J(\bm{v}^{[i+1]}).
        \end{eqnarray}
        Summing both sides of~\eqref{eq: sub_primal} over $i =0,\ldots,t$, yielding
        \begin{eqnarray}
        \begin{aligned}
        0 \leq \sum_{i=0}^{t}\Delta G(\bm{v}^{[i+1]};\bm{v}^{[i]}) &\leq J(\bm{v}^{[0]}) - J(\bm{v}^{[t+1]})\\
        &\leq J(\bm{v}^{[0]}) - \tilde{J},
        \end{aligned}
        \end{eqnarray}
        where $\tilde{J}$ is the lower bound of $J(\cdot)$. Allowing $t \to \infty$, we have 
        \begin{eqnarray}
                \lim_{i\to\infty} \Delta G(\bm{v}^{[i+1]};\bm{v}^{[i]}) = 0.
        \end{eqnarray}
        This completes the second statement~$(\RNum{2})$.
        
        For the last statement~$(iii)$, inspired by the proof line presented in~\cite{lu2014proximal}. By~\eqref{eq: sub-optimality}, we have
        \begin{eqnarray}\label{eq: sub_optimality2}
        \bm{0} = \beta(\bm{v}^{[i+1]} - \bm{v}^{[i]}) + \widetilde{\bm{x}}^{[i+1]} +\bm{u}^{[i+1]},
        \end{eqnarray}
        where $\widetilde{\bm{x}}^{[i+1]} \in \partial\langle\bm{w}^{[i]},\|\bm{v}^{[i+1]}\|_{\mathcal{G}_2}\rangle$, $\bm{u}^{[i]} \in \mathcal{N}_{\mathcal{C}}(\bm{v}^{[i]})$. Taking a dot-product with $(\bm{v}^{[i+1]} - \bm{v}^{[i]})$ on both sides of~\eqref{eq: sub_optimality2} yields
        \begin{eqnarray}
        \begin{aligned}
        \bm{0} = &\beta\Vert \bm{v}^{[i+1]} - \bm{v}^{[i]} \Vert_{2}^{2} + \langle\widetilde{\bm{x}}^{[i+1]}, \bm{v}^{[i+1]} - \bm{v}^{[i]}\rangle\\
        &+ \langle \bm{u}^{[i+1]}, \bm{v}^{[i+1]} - \bm{v}^{[i]}\rangle.
        \end{aligned}
        \end{eqnarray}
        By the definition of subgradient of the convex function, we have, 
        \begin{eqnarray}\label{eq: convex_def}
        \begin{aligned}
        \langle\bm{w}^{[i]},\|\bm{v}^{[i]}\|_{\mathcal{G}_2} - \|\bm{v}^{[i+1]}\|_{\mathcal{G}_2}\rangle &\geq \langle\widetilde{\bm{x}}^{[i+1]}, \bm{v}^{[i]} - \bm{v}^{[i+1]}\rangle\\
        &=\langle \bm{u}^{[i+1]}, \bm{v}^{[i+1]} - \bm{v}^{[i]}\rangle\\
        &\quad + \beta\Vert \bm{v}^{[i+1]} - \bm{v}^{[i]} \Vert_{2}^{2}.
        \end{aligned}
        \end{eqnarray}
        Therefore
        \begin{eqnarray}\label{eq: sub_decrease}
        \begin{aligned}
        &\quad\,\,G(\bm{v}^{[i]};\bm{v}^{[i]}) - G(\bm{v}^{[i+1]};\bm{v}^{[i]})\\
        &=\!\sum_{n=1}^{N}\!\sum_{k=1}^{K}w_{nk}^{[i]}(\|\bm{v}_{nk}^{[i]}\|_{2} - \|\bm{v}_{nk}^{[i+1]}\|_{2}) \!-\! \frac{\beta}{2}\|\bm{v}^{[i+1]}-\bm{v}^{[i]}\|_{2}^{2}\\
        &=\langle\bm{w}^{[i]},\|\bm{v}^{[i]}\|_{2} - \|\bm{v}^{[i+1]}\|_{2}\rangle - \frac{\beta}{2}\|\bm{v}^{[i+1]}-\bm{v}^{[i]}\|_{2}^{2}\\
        &\geq\langle \bm{u}^{[i+1]}, \bm{v}^{[i+1]} - \bm{v}^{[i]}\rangle + \frac{\beta}{2}\|\bm{v}^{[i+1]}-\bm{v}^{[i]}\|_{2}^{2}\\
        &= \frac{\beta}{2}\|\bm{v}^{[i]}-\bm{v}^{[i+1]}\|_{2}^{2} - \langle \bm{u}^{[i+1]}, \bm{v}^{[i]} - \bm{v}^{[i+1]}\rangle\\
        &\geq\frac{\beta}{2}\|\bm{v}^{[i]}-\bm{v}^{[i+1]}\|_{2}^{2},
        \end{aligned}
        \end{eqnarray}
        where the first inequality is obtained by~\eqref{eq: convex_def} and the last inequality holds since $\bm{u}^{[i+1]} \in \mathcal{N}_{\mathcal{C}}(\bm{v}^{[i+1]})$, completing the proof.
\section{Proof of Theorem~\ref{theorem: main theorem1}}\label{app_C}
        By Lemma~\ref{lem: lemma3}, the sequence $\{J(\bm{v}^{[i]})\}$ is monotonically decreasing. Since $J(\cdot)$ is coercive, we conclude that the sequence $\{\bm{v}^{[i]}\}$ is bounded. We conclude that $\{\bm{v}^{[i]}\}$ must have cluster points. Let $\bm{v}^{*}$ be a cluster point of $\{\bm{v}^{[i]}\}$. From Lemma~\ref{lem: Lemma2}, it suffices to show that $\bm{v}^{*} = \argmin_{\bm{v}}G(\bm{v};\bm{v}^{*})$. We prove this by contradiction. Assume that there exists a point $\tilde{\bm{v}}$ such that $\epsilon := G(\bm{v}^{*};\bm{v}^{*}) - G(\tilde{\bm{v}};\bm{v}^{*})>0$. Suppose $\{\bm{v}^{[i]}\}_{\mathcal{S}}\to \bm{v}^{*}$, $\mathcal{S} \subset \mathbb{N}$. Based on Lemma~\ref{lem: lemma3}, there must exist $k_1 > 0$ such that for all $k > k_1, k\in \mathcal{S}$ 
        \begin{eqnarray}\label{eq: assumption}
        G(\bm{v}^{[i]};\bm{v}^{[i]}) - G(\bm{v}^{[i+1]};\bm{v}^{[i]}) \leq \epsilon/4.
        \end{eqnarray}
        Note that $\bm{v}_{nk}^{[i]}\stackrel{\mathcal{S}}{\to}\bm{v}_{nk}^{*}$ and $w_{nk}^{[i]}\stackrel{\mathcal{S}}{\to}w_{nk}^{*}$, there exists $k_{2}$ such that for all $k>k_{2}, k\in \mathcal{S}$,
        $\!\sum_{n=1}^{N}\!\sum_{k=1}^{K}\!w_{nk}^{[i]}\Vert \bm{v}_{nk}^{[i]}\Vert_{2} - w_{nk}^{*}\Vert\bm{v}_{nk}^{*}\Vert_{2 }>-\epsilon/10,$ and $
        \frac{\beta}{2}\Vert\bm{v}^{*} \!-\!\tilde{\bm{v}}\Vert_{2}^{2} + \!\sum_{n,k}\!(w_{nk}^{*}\!-\!w_{nk}^{[i]})\Vert\tilde{\bm{v}}_{nk}\Vert_{2} \!-\!\frac{\beta}{2}\Vert\tilde{\bm{v}}\! -\! \bm{v}^{[i]}\Vert_{2}^{2} >-\epsilon/10$.

        Therefore, for all $k>k_{2}, k \in \mathcal{S}$,
        \begin{eqnarray}
        \begin{aligned}
        &\quad\,\,G(\bm{v}^{*};\bm{v}^{*}) - G(\tilde{\bm{v}};\bm{v}^{[i]})\\
        &= \sum_{n=1}^{N}\sum_{k=1}^{K}w_{nk}^{*}\Vert \bm{v}_{nk}^{*}\Vert_{2} - \frac{\beta}{2}\Vert\tilde{\bm{v}} - \bm{v}^{[i]}\Vert_{2}^{2}\\ 
        &\quad- \sum_{n=1}^{N}\sum_{k=1}^{K}(w_{nk}^{*}-(w_{nk}^{*}-w_{nk}^{[i]}))\Vert \tilde{\bm{v}}_{nk}\Vert_{2}\\
        &= \left[G(\bm{v}^{*};\bm{v}^{*}) - G(\tilde{\bm{v}};\bm{v}^{*})\right]+\frac{\beta}{2}\Vert\tilde{\bm{v}} -\bm{v}^{*}\Vert_{2}^{2}\\
        & \quad+ \sum_{n=1}^{N}\sum_{k=1}^{K}(w_{nk}^{*} - w_{nk}^{[i]})\Vert\tilde{\bm{v}}_{nk}\Vert_{2} - \frac{\beta}{2}\Vert\tilde{\bm{v}} - \bm{v}^{[i]}\Vert_{2}^{2}\\
        &\geq \epsilon - \epsilon/10 = 9\epsilon/10,
        \end{aligned}
        \end{eqnarray}
        and that
        \begin{eqnarray}
        \begin{aligned}
        &\quad\,\, G(\bm{v}^{[i]};\bm{v}^{[i]}) - G(\bm{v}^{*};\bm{v}^{*})\\
        &= \sum_{n=1}^{N}\sum_{k=1}^{K}w_{nk}^{[i]}\Vert \bm{v}_{nk}^{[i]}\Vert_{2} - \sum_{n=1}^{N}\sum_{k=1}^{K}w_{nk}^{*}\Vert \bm{v}_{nk}^{*}\Vert> -\epsilon/10.
        \end{aligned}
        \end{eqnarray}
        Hence, for all $k>\max(k_1,k_2), k\in\mathcal{S}$, it holds that
        \begin{eqnarray}
        \begin{aligned}
        &\quad\,\,G(\bm{v}^{[i]};\bm{v}^{[i]}) - G(\tilde{\bm{v}};\bm{v}^{[i]})\\
        &=  G(\bm{v}^{[i]};\bm{v}^{[i]})  - G(\bm{v}^{*};\bm{v}^{*}) + G(\bm{v}^{*};\bm{v}^{*}) - G(\tilde{\bm{v}};\bm{v}^{[i]})\\
        &\geq -\epsilon/10 + 9\epsilon/10 = 4\epsilon/5,
        \end{aligned}
        \end{eqnarray}
        contradicting~\eqref{eq: assumption}. Hence, $\bm{v}^{*} = \argmin_{\bm{v}}G(\bm{v};\bm{v}^{*})$. By Lemma~\ref{lem: Lemma2}, $\bm{v}^{*}$ satisfies the first-order necessary optimality for~\eqref{eq:unconstrainted}. This completes the proof.
\section{Proof of Lemma~\ref{lem: Optimality residual is bounded}}\label{app_D}
        Recall the first-order necessary optimality condition~\eqref{eq: sub-optimality} of subproblem~\eqref{eq: convex_subproblem}. By rearranging the term, and we have 
        \begin{eqnarray}\label{eq:sub-opt-rearrange}
        \bm{v}^{[i]} - \bm{v}^{[i+1]} = \frac{1}{\beta}(\bm{w}^{[i]}\odot\bm{x}^{[i+1]} + \bm{u}^{[i+1]}),
        \end{eqnarray}
        where $\bm{x}^{[i+1]} \in \partial\|\bm{v}^{[i+1]}\|_{\mathcal{G}_2}$ and $\bm{u}^{[i+1]} \in \mathcal{N}_{\mathcal{C}}(\bm{v}^{[i+1]})$. Square on both sides of~\eqref{eq:sub-opt-rearrange}, we have
        \begin{eqnarray}
        \begin{aligned}
        &\quad\,\,\|\bm{v}^{[i]} - \bm{v}^{[i+1]}\|_{2}^{2}\\
        &=      \frac{1}{\beta^{2}}\|\bm{w}^{[i]}\odot\bm{x}^{[i+1]} + \bm{u}^{[i+1]}\|_{2}^{2} \\
        &= \frac{1}{\beta^{2}}\|(\bm{w}^{[i]} - \bm{w}^{[i+1]})\odot\bm{x}^{[i+1]} + \bm{r}^{[i+1]}\|_{2}^{2}\\
        &=\frac{1}{\beta^{2}}\|(\bm{w}^{[i]} - \bm{w}^{[i+1]})\odot\bm{x}^{[i+1]} \|_{2}^{2} +\frac{1}{\beta^{2}}\|\bm{r}^{[i+1]}\|_{2}^{2}\\
        & \quad+ \frac{2}{\beta^{2}}\la(\bm{w}^{[i]} - \bm{w}^{[i+1]})\odot\bm{x}^{[i+1]}, \bm{r}^{[i+1]}\ra\\
        &=\frac{1}{\beta^{2}}\|(\bm{w}^{[i]} - \bm{w}^{[i+1]})\odot\bm{x}^{[i+1]} \|_{2}^{2} +\frac{1}{\beta^{2}}\|\bm{r}^{[i+1]}\|_{2}^{2}\\
        & \quad- \frac{2}{\beta^2}\la(\bm{w}^{[i]} - \bm{w}^{[i+1]})\odot\bm{x}^{[i+1]},\\
        &\quad (\bm{w}^{[i]}-\bm{w}^{[i+1]})\odot\bm{x}^{[i+1]}+\beta(\bm{v}^{[i+1]} - \bm{v}^{[i]})\ra\\
        &= -\frac{1}{\beta^{2}}\|(\bm{w}^{[i]} - \bm{w}^{[i+1]})\odot\bm{x}^{[i+1]} \|_{2}^{2}+\frac{1}{\beta^{2}}\|\bm{r}^{[i+1]}\|_{2}^{2}\\
        &\quad -\frac{2}{\beta}\la(\bm{w}^{[i]} - \bm{w}^{[i+1]})\odot\bm{x}^{[i+1]},\bm{v}^{[i+1]} - \bm{v}^{[i]}\ra,
        \end{aligned}
        \end{eqnarray}
        where we exploit $\bm{u}^{[i+1]} = \beta(\bm{v}^{[i]}-\bm{v}^{[i+1]})-\bm{w}^{[i]}\odot\bm{x}^{[i+1]}$ to obtain the fourth equality. Then we have
        \begin{eqnarray}
        \begin{aligned}
        &\quad\,\,\|\bm{r}^{[i+1]}\|_{2}^{2}\\
        &= \beta^2\|\bm{v}^{[i]} - \bm{v}^{[i+1]}\|_{2}^{2} + \|(\bm{w}^{[i]} - \bm{w}^{[i+1]})\odot\bm{x}^{[i+1]} \|_{2}^{2}\\
        &\quad+2\beta\la(\bm{w}^{[i]}-\bm{w}^{[i+1]})\odot\bm{x}^{[i+1]},\bm{v}^{[i+1]}-\bm{v}^{[i]}\ra\\
        &\leq \beta^2\|\bm{v}^{[i]} - \bm{v}^{[i+1]}\|_{2}^{2} + \|\bm{w}^{[i]} - \bm{w}^{[i+1]}\|_{2}^{2}\\
        &\quad+ 2\beta \|(\bm{w}^{[i]}-\bm{w}^{[i+1]})\odot\bm{x}^{[i+1]}\|_{2}\|\bm{v}^{[i]}-\bm{v}^{[i+1]}\|_{2}\\
        &\leq \beta^2\|\bm{v}^{[i]}-\bm{v}^{[i+1]}\|_{2}^{2} +\|\bm{w}^{[i]} - \bm{w}^{[i+1]}\|_{2}^{2}\\
        & \quad+ 2\beta\|\bm{w}^{[i]} - \bm{w}^{[i+1]}\|_{2}\|\bm{v}^{[i]}-\bm{v}^{[i+1]}\|_{2}\\
        &\leq \beta^2\|\bm{v}^{[i]}-\bm{v}^{[i+1]}\|_{2}^{2} + p^4\kappa^{2}\|\bm{v}^{[i]}-\bm{v}^{[i+1]}\|_{2}^{2}\\
        & \quad+ 2\beta p^2\kappa\|\bm{v}^{[i]}-\bm{v}^{[i+1]}\|_{2}^{2}\\
        &= (\beta^2 +2\beta \kappa p^2 +\kappa^{2}p^4)\|\bm{v}^{[i]}-\bm{v}^{[i+1]}\|_{2}^{2},
        \end{aligned}
        \end{eqnarray}
        where the last inequality holds due to $f'_{p}(\cdot)$ is $p^{2}$-Lipschitz continuous and $\kappa = \rho^{\max}$. This completes the proof.

\section{Construction of Fenchel-Rockafellar Dual~\eqref{eq: sub-dual1}}\label{app_A0}
This construction of the dual program owes to the conjugate function~\cite{rockafellar2015convex}. We first rewrite the objective $G(\bm{v};\bm{v}^{[i]})$ by exploiting the conjugate function as
\begin{eqnarray}\label{eq: same_pri_sub}
\begin{aligned}
G(\bm{v};\bm{v}^{[i]}) &= \sum_{n=1}^{N}\sum_{k=1}^{K}w_{nk}^{[i]}[\sup_{\bm{\lambda}_{nk}}\la\bm{\lambda}_{nk},\bm{v}_{nk}\ra-\delta_{\mathbb{B}}(\bm{\lambda}_{nk})]\\
&\quad+\frac{\beta}{2}\|\bm{v}-\bm{v}^{[i]}\|_{2}^{2} + \sup_{\bm{\mu}}[\la\bm{\mu},\bm{v}\ra-\delta_{\mathcal{C}}^{*}(\bm{\mu})],
\end{aligned}
\end{eqnarray}
where $\mathbb{B} := \{\bm{y}\in \mathbb{C}^{L}\mid\|\bm{y}\|_{2}\leq 1\}$ denotes the dual norm unit ball of $\|\cdot\|_{2}$. Then the primal subproblem~\eqref{eq: unconstrained_sub} is simply
\begin{equation}\label{eq: reweite_Pri_sub}
\begin{aligned}
\inf_{\bm{v}}G(\bm{v};\bm{v}^{[i]}) =   \inf_{\bm{v}}\sup_{\bm{\lambda},\bm{\mu}}~&\sum_{n=1}^{N}\sum_{k=1}^{K}w_{nk}^{[i]}[\la\bm{\lambda}_{nk},\bm{\mu}_{nk}\ra-\delta_{\mathbb{B}}(\bm{\lambda}_{nk})]\\
&+\frac{\beta}{2}\|\bm{v}-\bm{v}^{[i]}\|_{2}^{2}+ \la\bm{\mu},\bm{v}\ra-\delta_{\mathcal{C}}^{*}(\bm{\mu}).
\end{aligned}
\end{equation}
Swapping the order of $\inf$ and $\sup$ gives the corresponding Lagrangian dual, i.e.,
\begin{eqnarray}\label{eq: dual2}
\begin{aligned}
\sup_{\bm{\lambda},\bm{\mu}}\inf_{\bm{v}}~&\sum_{n=1}^{N}\sum_{n=k}^{K}w_{nk}^{[i]}[\la\bm{\lambda}_{nk},\bm{v}_{nk}\ra-\delta_{\mathbb{B}}(\bm{\lambda}_{nk})]\\
&+\frac{\beta}{2}\|\bm{v}-\bm{v}^{[i]}\|_{2}^{2}+ \la\bm{\mu},\bm{v}\ra-\delta_{\mathcal{C}}^{*}(\bm{\mu}).
\end{aligned}
\end{eqnarray}
The dual objective at $\bm{v}^{[i]}$ is then given by
\begin{equation}\label{eq:dual_form}
\begin{aligned}
Q(\bm{\lambda},\bm{\mu};\bm{v}^{[i]}) = \inf_{\bm{v}}~&\sum_{n=1}^{N}\sum_{k=1}^{K}w_{nk}^{[i]}[\la\bm{\lambda}_{nk},\bm{v}_{nk}\ra-\delta_{\mathbb{B}}(\bm{\lambda}_{nk})]\\
&+\frac{\beta}{2}\|\bm{v}-\bm{v}^{[i]}\|_{2}^{2}+ \la\bm{\mu},\bm{v}\ra-\delta_{\mathcal{C}}^{*}(\bm{\mu}).
\end{aligned}
\end{equation}
By exploiting the first-order necessary optimality condition of~\eqref{eq:dual_form}, we have
\begin{equation}\label{eq:dual3}
\begin{aligned}
\sup_{\bm{\lambda},\bm{\mu}}~& -\frac{1}{2\beta}\|\bm{\lambda}\odot\bm{w}^{[i]}+\bm{\mu}-\beta\bm{v}^{[i]}\|_{2}^{2}+\frac{\beta}{2}\|\bm{v}^{[i]}\|_{2}^{2}-\delta_{\mathcal{C}}^{*}(\bm{\mu})\\
&\quad-\sum_{n=1}^{N}\sum_{k=1}^{K}w_{nk}^{[i]}\delta_{\mathbb{B}}(\bm{\lambda}_{nk}),
\end{aligned}
\end{equation}
which is consistent with the presented Fenchel-Rockafellar in~\eqref{eq: sub-dual1}. This completes the construction.
\section{Proof of Theorem~\ref{theo: duality gap gone}}\label{app_E}
        We evaluate $g(\bm{v},\bm{x},\bm{u};\bm{v}^{[i]})$ at $\bm{v}^{[i]}$ with $\bm{\lambda}^{[i+1]} \!\in\! \partial \|\bm{v}^{[i+1]}\|_{\mathcal{G}_2}$ and $\bm{\mu}^{[i+1]}\in\mathcal{N}_{\mathcal{C}}(\bm{v}^{[i+1]})$. Then, we have 
        \begin{equation}\label{eq:small g}
        \begin{aligned}
        &\quad\,\,g(\bm{v}^{[i]},\bm{\lambda}^{[i+1]},\bm{\mu}^{[i+1]};\bm{v}^{[i]})\\
        &= G(\bm{v}^{[i]};\bm{v}^{[i]})-  Q(\bm{\lambda}^{[i+1]},\bm{\mu}^{[i+1]};\bm{v}^{[i]})\\ 
        &=\sum_{n=1}^{N}\sum_{k=1}^{K}w_{nk}^{[i]}\Vert\bm{v}_{nk}^{[i]}\Vert_{2}+ \delta_{\mathcal{C}}(\bm{v}^{[i]})-\frac{\beta}{2}\|\bm{v}^{[i]}\|_{2}^{2}\\
        &\quad+\frac{1}{2\beta}\|\bm{\lambda}^{[i+1]}\odot\bm{w}^{[i]}+\bm{\mu}^{[i+1]}-\beta\bm{v}^{[i]}\|_{2}^{2}+\delta_{\mathcal{C}}^{*}(\bm{\mu}^{[i+1]})\\
        &\geq\frac{1}{2\beta}\Vert\bm{w}^{[i]}\!\odot\!\bm{\lambda}^{[i+1]}\! +\! \bm{\mu}^{[i+1]}\Vert_{2}^{2}\!+\!\la\bm{w}^{[i]},\|\bm{v}^{[i]}\| - \bm{v}^{[i]}\!\odot\!\bm{\lambda}^{[i+1]}\ra\\
        &\quad +\la\bm{\mu}^{[i+1]},\bm{v}^{[i]}-\bm{v}^{[i]}\ra\\
        &\geq \frac{1}{2\beta}\Vert\bm{w}^{[i]}\odot\bm{\lambda}^{[i+1]} + \bm{\mu}^{[i+1]}\Vert_{2}^{2}, \\
        \end{aligned}
        \end{equation}
        where the first inequality holds due to the Fenchel-Young inequality and the second inequality is obtained since $\Vert\bm{\lambda}^{[i+1]}\Vert_{2}\leq 1$ makes $\|\bm{v}^{[i]}\|_{\mathcal{G}_2} - \bm{v}^{[i]}\odot\bm{\lambda}^{[i+1]} \geq 0$.
        
        Next, by making use of~\eqref{eq: sub-optimality} to replace $\bm{\mu}^{[i+1]}$ with $\beta(\bm{v}^{[i]}-\bm{v}^{[i+1]}) - \bm{w}^{[i]}\odot \bm{\lambda}^{[i+1]}$ in~\eqref{eq:small g}, we have 
        
        \begin{equation}\label{eq: houxu}
        \begin{aligned}
        &\quad\,\,\frac{1}{2\beta}\Vert\bm{w}^{[i]}\odot\bm{\lambda}^{[i+1]} + \bm{\mu}^{[i+1]}\Vert_{2}^{2}\\
        &= \frac{1}{2\beta}\Vert\bm{w}^{[i+1]}\!\odot\!\bm{\lambda}^{[i+1]} + \bm{\mu}^{[i+1]}+(\bm{w}^{[i]} - \bm{w}^{[i+1]})\!\odot\!\bm{\lambda}^{[i+1]}\Vert_{2}^{2}\\
        &=\frac{1}{2\beta}( \Vert\bm{r}^{[i+1]}\Vert_{2}^{2} + \Vert(\bm{w}^{[i]} - \bm{w}^{[i+1]})\odot\bm{\lambda}^{[i+1]}\Vert_{2}^{2}\\
        &\quad + 2\la\bm{r}^{[i+1]}, (\bm{w}^{[i]} - \bm{w}^{[i+1]})\odot\bm{\lambda}^{[i+1]}\ra) \\
        &= \frac{1}{2\beta}\Vert\bm{r}^{[i+1]}\Vert_{2}^{2}- \frac{1}{2\beta}\Vert(\bm{w}^{[i]} - \bm{w}^{[i+1]})\odot\bm{\lambda}^{[i+1]}\Vert_{2}^{2}\\
        &\quad-\la\bm{v}^{[i+1]}-\bm{v}^{[i]},(\bm{w}^{[i]} - \bm{w}^{[i+1]})\odot\bm{\lambda}^{[i+1]}\ra\\
        &\geq\frac{1}{2\beta}\Vert\bm{r}^{[i+1]}\Vert_{2}^{2}-\Vert\bm{v}^{[i]}-\bm{v}^{[i+1]}\Vert_{2}\Vert\bm{w}^{[i]} - \bm{w}^{[i+1]}\Vert_{2}\\
        &\quad-\frac{1}{2\beta}\Vert\bm{w}^{[i]}-\bm{w}^{[i+1]}\Vert_{2}^{2},
        \end{aligned}
        \end{equation}
        where we exploit $\text{Cauchy\textendash Bunyakovskii\textendash Schwarz}$ (CBS) inequality to obtain the last inequality.
        
        On the other hand, by strong duality, we have 
        \begin{equation}\label{eq:strong}
        \begin{aligned}
        g(\bm{v}^{[i]},\bm{\lambda}^{[i+1]},\bm{\mu}^{[i+1]};\bm{v}^{[i]}) \!=\! G(\bm{v}^{[i]};\bm{v}^{[i]}) - G(\bm{v}^{[i+1]};\bm{v}^{[i]}).
        \end{aligned}
        \end{equation}
        Combine~\eqref{eq:small g},~\eqref{eq: houxu} and~\eqref{eq:strong}, we have
        \begin{equation}
        \begin{aligned}
        &\quad\,\,\Vert\bm{r}^{[i+1]}\Vert_{2}^{2}\\
        &\leq 2\beta\Delta G(\bm{v}^{[i+1]};\bm{v}^{[i]})
        +\Vert\bm{w}^{[i]} - \bm{w}^{[i+1]}\Vert_{2}^{2}\\
        & \quad+ 2\beta\Vert\bm{v}^{[i]}-\bm{v}^{[i+1]}\Vert_{2}\Vert\bm{w}^{[i]} - \bm{w}^{[i+1]}\Vert_{2}\\
        &\leq  2\beta\Delta G(\bm{v}^{[i+1]};\bm{v}^{[i]}) + \kappa^{2}p^{4}\Vert\bm{v}^{[i]} - \bm{v}^{[i+1]}\Vert_{2}^{2}\\
        &\quad + 2\beta \kappa p^{2}\Vert\bm{v}^{[i]}-\bm{v}^{[i+1]}\Vert_{2}^{2}\\
        &\leq 2\beta\Delta G(\bm{v}^{[i+1]};\bm{v}^{[i]}) +(\kappa^{2}p^{4} + 2\beta \kappa p^{2})\Vert\bm{v}^{[i]}-\bm{v}^{[i+1]}\Vert_{2}^{2}\\
        &\leq 2\beta\Delta G(\bm{v}^{[i+1]};\bm{v}^{[i]}) + \frac{2}{\beta}(\kappa^{2}p^{4} + 2\beta \kappa p^{2})\Delta G(\bm{v}^{[i+1]};\bm{v}^{[i]})\\
        &\leq \frac{2}{\beta}(\beta^{2}+ 2\beta \kappa p^{2}+\kappa^{2}p^{4} )\Delta G(\bm{v}^{[i+1]};\bm{v}^{[i]})\\
        & \leq \frac{2}{\beta}(\beta^{2}+ 2\beta \kappa p^{2}+\kappa^{2}p^{4} )(J(\bm{v}^{[i]})-J(\bm{v}^{[i+1]})),\\
        \end{aligned}
        \end{equation}
        where the fourth inequality holds due to Lemma~\ref{lem: lemma3} $(\RNum{3})$.
        
        Summing up both sides of above inequality from $i=0$ to $t$, we have
        \begin{equation}
        \begin{aligned}
        C(J(\bm{v}^{[0]}) - J(\bm{v}^{[t+1]}))&\geq \sum_{i=0}^{t}\Vert\bm{r}^{[i+1]}\Vert_{2}^{2} \\
        &\geq t\min_{i=0,\cdots,t}\Vert\bm{r}^{[i+1]}\Vert_{2}^{2}
        \end{aligned}
        \end{equation}
        with $C= \frac{2(\beta^{2}+ 2\beta \kappa p^{2}+\kappa^{2}p^{4})}{\beta}$. This indicates that 
        \begin{equation}\label{eq: frac 1/t}
        \begin{aligned}
        \min_{i=0,\cdots,t} \Vert\bm{r}^{[i+1]}\Vert_{2}^{2}&\leq\! \frac{2(\beta^{2}+ 2\beta \kappa p^{2}+\kappa^{2}p^{4})}{t\beta}(J(\bm{v}^{[0]})\! -\! J(\bm{v}^{[t+1]}))\\
        &\leq \frac{2(\beta^{2}+ 2\beta \kappa p^{2}+\kappa^{2}p^{4})}{t\beta}(J(\bm{v}^{[0]}) - J(\bm{v}^{*})).
        \end{aligned}
        \end{equation}
        Hence
        \begin{equation}
        \min_{i=0,\cdots,t} \Vert\bm{r}^{[i+1]}\Vert_{2}^{2} = O(\frac{1}{t}).
        \end{equation}
        This completes the proof.

\bibliographystyle{ieeetr}
\bibliography{reference}

\end{document}